\PassOptionsToPackage{sort&compress}{natbib}
\documentclass[aps,reprint,10pt,prx,superscriptaddress,twocolumn,longbibliography,nofootinbib,floatfix]{revtex4-2}
\usepackage[T1]{fontenc}
\usepackage[utf8]{inputenc}
\usepackage{amsmath,amssymb,bm,xcolor,graphicx,booktabs}
\usepackage{amsthm}
\usepackage{xspace}
\usepackage{tikz}
\usetikzlibrary{arrows.meta}
\usepackage[colorlinks=true,linkcolor=teal,citecolor=teal,urlcolor=teal,pdfpagemode=UseNone]{hyperref}
\renewcommand{\figurename}{Fig.}
\renewcommand{\tablename}{Table}
\makeatletter
\renewcommand{\fnum@figure}{\textbf{\figurename~\thefigure}}
\renewcommand{\fnum@table}{\textbf{\tablename~\thetable}}
\makeatother
\newcommand{\cBorn}{\cite{herrero-gonzalez_born_2025a}\xspace}
\newcommand{\cTonner}{\cite{tonner_parametric_2026}\xspace}
\newcommand{\cZhou}{\cite{zhou_autonomous_2026}\xspace}
\newcommand{\cArmengol}{\cite{recio-armengol_train_2026a}\xspace}

\newcommand{\cGretton}{\cite{gretton_kernel_2012}\xspace}

\newcommand{\cPerez}{\cite{perez_film_2017}\xspace}
\newcommand{\cKingma}{\cite{kingma_adam_2017}\xspace}

\newcommand{\cRende}{\cite{rende_foundation_2025}\xspace}

\newcommand{\cCong}{\cite{cong_quantum_2019}\xspace}
\newcommand{\cFuchs}{\cite{fuchs_cryptographic_1998}\xspace}
\newcommand{\cTagliacozzo}{\cite{tagliacozzo_scaling_2008}\xspace}
\newcommand{\cTao}{\cite{tao_discriminative_2024}\xspace}

\newcommand{\cFishman}{\cite{fishman_itensor_2021}\xspace}
\newcommand{\cOrus}{\cite{orus_practical_2014a}\xspace}

\newcommand{\cFerris}{\cite{ferris_perfect_2012}\xspace}

\newcommand{\cHuangShadow}{\cite{huang_predicting_2020}\xspace}

\newcommand{\cWetzel}{\cite{wetzel_unsupervised_2017}\xspace}

\newcommand{\cCoyleBorn}{\cite{coyle_born_2020}\xspace}
\newcommand{\cCoyleFinance}{\cite{coyle_quantum_2020a}\xspace}
\newcommand{\cDias}{\cite{dias_classical_2024}\xspace}
\newcommand{\cKiss}{\cite{kiss_conditional_2022}\xspace}

\newcommand{\la}{{\boldsymbol{\lambda}}}
\newcommand{\tht}{{\boldsymbol{\theta}}}
\newcommand{\phiv}{{\boldsymbol{\phi}}}
\newcommand{\etav}{{\boldsymbol{\eta}}}
\newcommand{\E}{\mathbb E}
\newcommand{\Var}{\operatorname{Var}}

\newcommand{\KL}{D_{\rm KL}}
\newcommand{\MMD}{\operatorname{MMD}}

\newtheorem{proposition}{Proposition}
\newtheorem*{informalproposition}{Proposition (informal)}
\newtheorem{corollary}{Corollary}
\begin{document}
\title{Classically Trained Quantum Generative Models with Shared Training and Fine Tuning for Quantum Matter}
\author{Mario Herrero-Gonzalez}
\email{mario.herrero@ed.ac.uk}
\affiliation{School of Informatics, University of Edinburgh, 10 Crichton Street, Edinburgh EH8 9AB, United Kingdom}
\author{Elham Kashefi}
\affiliation{School of Informatics, University of Edinburgh, 10 Crichton Street, Edinburgh EH8 9AB, United Kingdom}
\affiliation{LIP6, CNRS, Sorbonne Universit\'e, 4 place Jussieu, 75005 Paris, France}
\author{Mina Doosti}
\affiliation{School of Informatics, University of Edinburgh, 10 Crichton Street, Edinburgh EH8 9AB, United Kingdom}
\begin{abstract}
Ground-state measurement statistics characterize quantum matter. Selected correlations can guide the classical training of a quantum sampler even when sampling the full distribution is difficult. We ask whether training across related ground-state distributions helps the sampler reproduce their correlations more accurately. We train a hypernetwork to map Hamiltonian parameters to circuit angles, sharing information across the phase diagram before local fine-tuning with additional reference data for quantum deployment. For classical training, we derive an unbiased Monte Carlo estimator of Fourier coefficients for circuits built from general Pauli rotations, with a cost that depends on the circuit generators. We also introduce a CDF-KL loss for comparing selected marginals. We study the Haldane Hamiltonian and transverse-field Ising model, using an automated search to select reference calculations near phase transitions. On the twelve-site chain, the adaptively pretrained hypernetwork matches the independent fits' final mean total variation distance in fewer local updates and continues to a lower error. On a 1,024-site chain and a 4 by 256 lattice, the fine-tuned circuits have lower squared maximum mean discrepancy from the reference distributions than the independent circuits at the evaluated point in each geometry. These results motivate shared training for preparing quantum samplers across a phase diagram and local fine-tuning.
\end{abstract}

\maketitle

\section{Introduction}
\vspace{-3pt}
Quantum computers are natural samplers, measurements draw outcomes from the Born distribution. Quantum circuit Born machines (QCBMs) learn target distributions by training circuit angles on reference data~\cite{liu_differentiable_2018,benedetti_generative_2019,coyle_born_2020,bakó_probleminformed_2026}. Their potential advantage requires distributions that are hard to sample classically, yet can be trained to reproduce relevant data.

Complexity results for instantaneous quantum polynomial-time (IQP) circuits and random circuit sampling support classical sampling hardness under specified assumptions and error criteria~\cite{bremner_averagecase_2016,bouland_complexity_2019,bremner_achieving_2017}. Both have been implemented experimentally~\cite{arute_quantum_2019,bluvstein_logical_2024}, but sampling hardness does not establish efficient training on a specified dataset.

Two complementary approaches address training. The first develops practical objectives, gradient estimators and initialization strategies, including training Ising Born machines~\cCoyleBorn. IQP models can be trained classically using maximum mean discrepancy (MMD) and deployed as quantum samplers~\cite{armengol_iqpopt_2026,recio-armengol_train_2026a}. Classically accessible correlators provide the loss and its gradients without requiring efficient sampling of the full model distribution~\cite{recio-armengol_train_2026a,bakó_fermionic_2025,kolarovszki_generative_2026,kurkin_universality_2026}.

The second studies trainability and classical simulation. Barren plateaus have exponentially small gradient variances~\cite{mcclean_barren_2018} and depend on the circuit, loss and expressibility~\cite{cerezo_cost_2021,holmes_connecting_2022}. Many provably plateau-free models admit classical simulation, sometimes after quantum data acquisition~\cite{cerezo_does_2025}. Certain anticoncentrated generative ensembles exhibit loss concentration, while restrictions that improve trainability can permit classical sampling~\cite{herbst_limits_2025}. Trainability does not necessarily imply dequantizability~\cite{gil-fuster_relation_2025}.

\begin{figure}[!t]
\centering
\setlength{\abovecaptionskip}{1pt}
\begin{tikzpicture}
  \node[anchor=south west,inner sep=0] (artwork) at (0,0)
    {\includegraphics[trim=60bp 8bp 109.8bp 0,clip,width=\columnwidth]{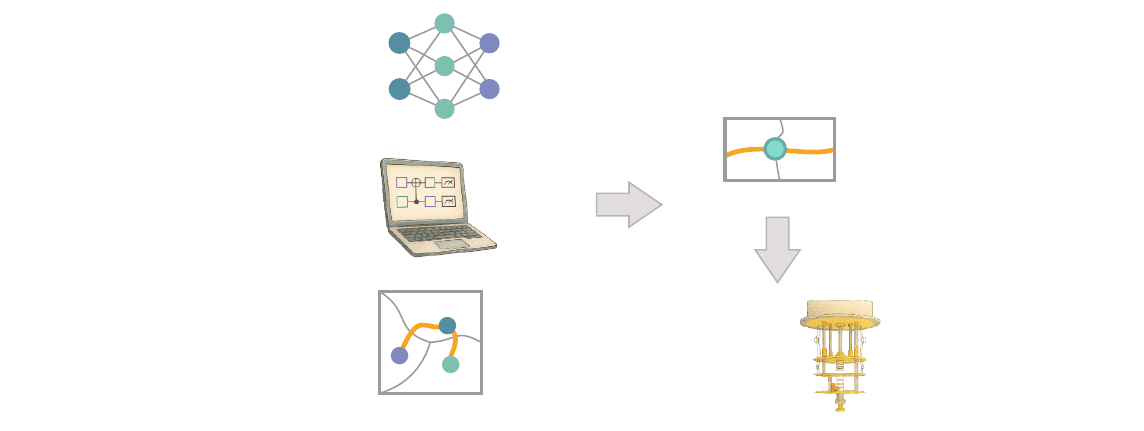}};
  \begin{scope}[x=1.44778481\columnwidth,y=0.54489715\columnwidth,
    xshift=-0.15822785\columnwidth,yshift=-0.02109705\columnwidth,
    every node/.style={inner sep=0,align=center,font=\fontsize{8}{9}\selectfont}]
    \node at (0.218,0.869) {Hypernetwork};
    \node at (0.218,0.791) {Sec.~\ref{sec:family}};
    \node at (0.325,0.834) {$\boldsymbol\lambda$};
    \node at (0.464,0.909) {$\boldsymbol\gamma_\phiv$};
    \node at (0.464,0.788) {$\boldsymbol\beta_\phiv$};

    \node at (0.218,0.570) {Classical Surrogate};
    \node at (0.218,0.493) {Training, Sec.~\ref{sec:training}};
    \node at (0.465,0.630)
      {$\boldsymbol\theta(\boldsymbol\gamma_\phiv,\boldsymbol\beta_\phiv)$};

    \node at (0.218,0.287) {Autonomous};
    \node at (0.218,0.211) {Phase Discovery};
    \node at (0.218,0.134) {Sec.~\ref{sec:acquisition}};
    \node[anchor=west] at (0.429,0.320) {New $\boldsymbol\lambda$};
    \node[anchor=west] at (0.429,0.184) {No labels};
    \node[anchor=west] at (0.429,0.106) {No sweep};

    \node at (0.682,0.925) {Optional};
    \node at (0.682,0.849) {Fine Tuning};
    \node at (0.682,0.773) {Sec.~\ref{sec:sharing}};
    \node at (0.643,0.219) {Quantum};
    \node at (0.643,0.143) {Deployment};
  \end{scope}
\end{tikzpicture}
\caption{\textbf{Learning across a phase diagram to sample at selected Hamiltonian points $\la$.} The hypernetwork maps $\la$ to scale and shift vectors that set the circuit angles with shared base angles, Eq.~\eqref{eq:hyper}. Both are trained jointly across reference distributions. Autonomous discovery adds reference-state calculations near rapid distribution changes, without phase labels or a dense sweep. Local fine-tuning uses additional reference data before quantum deployment.}
\label{fig:workflow}
\vspace{-10pt}
\end{figure}

For IQP models, MMD trainability depends on circuit generators and kernel spectrum~\cite{shen_characterizing_2026}, while data-informed initialization can improve convergence~\cite{lerch_iqp_2026}. Different QCBMs can agree on the correlations used for training and differ on other features of the distribution~\cBorn.

Neural representations can learn many-body ground states and their measurement statistics~\cite{carleo_solving_2017,torlai_manybody_2018}. This motivates the following question.
\begin{center}
\textit{Can learning from related ground-state distributions improve a classically trained quantum sampler?}
\end{center}

We use reference ground-state data to learn quantum samplers across the phase diagram and fine-tune them at selected points. To this end, we jointly train a hypernetwork and shared QCBM parameters across a family of Hamiltonians. The hypernetwork maps Hamiltonian parameters to circuit angles through a scale-and-shift construction~\cTonner, while each update combines losses from several reference distributions. Learning across Hamiltonians also appears in neural quantum states~\cRende and in provable prediction of ground-state properties within a quantum phase~\cite{huang_provably_2022,lewis_improved_2024}. A related approach introduces the external parameters directly into the Born-machine circuit through input gates~\cKiss. We test whether shared training helps recover correlations absent from the loss.

Obtaining reference ground states can be costly, so we also use the learned family to choose where to request additional data. Rapid changes in its measurement distribution guide this search towards possible phase boundaries, without requiring a predefined order parameter.

Our starting point is likelihood training, equivalent to minimizing an empirical Kullback--Leibler (KL) divergence objective across the reference distributions. When evaluating model probabilities becomes costly, we use selected Fourier coefficients instead. As an alternative to MMD, we use KL divergences to compare the cumulative distribution functions (CDFs) of selected marginals. We call this loss CDF-KL. We also derive an unbiased Monte Carlo (MC) estimator for the model coefficients of ordered, generally noncommuting Pauli circuits. Its cost depends on the number of amplitudes evolved in each MC sample, which is determined by the selected gates. The choice of generators therefore determines both the cost of classical training and the circuit's inductive bias, including its preserved Pauli symmetries.

Using Monte Carlo coefficients also affects how we measure agreement with the target distribution. Sampling fluctuations can bias a nonlinear loss even when each coefficient estimate is unbiased. We correct this bias for squared MMD~\cite{gretton_kernel_2012,recio-armengol_train_2026a}, while our sampled CDF-KL loss remains biased. The sampled MMD curves in Figs.~\ref{fig:large} and~\ref{fig:large-2d} use the variance correction derived in Appendix~\ref{app:mmd-evaluation}.

Together, shared training, classical coefficient estimation and local fine-tuning provide a framework for learning quantum samplers across a phase diagram and testing whether sharing information across Hamiltonians improves the fit to individual target distributions.

The paper is organized as follows. Sec.~\ref{sec:family} introduces the target distributions, hypernetwork and circuit generators. Sec.~\ref{sec:training} develops classical training with selected Fourier coefficients, including the Monte Carlo estimator and the CDF-KL and MMD losses. In Sec.~\ref{sec:results}, we examine distribution and response accuracy in small transverse-field Ising and Haldane systems, then restrict the reference data, coefficient information and allowed gates. In Sec.~\ref{sec:acquisition}, we adapt the autonomous search of Ref.~\cZhou to concentrate costly reference calculations near phase boundaries. Sec.~\ref{sec:sharing} compares shared and independent training at common boundary points and examines local fine-tuning. Sec.~\ref{sec:large-systems} describes the large-system protocol using selected-marginal responses. Sec.~\ref{sec:discussion} discusses the findings and limitations. Figure~\ref{fig:workflow} summarizes the training and deployment procedure. The appendices provide derivations and numerical settings.

\section{Learning a family of measurement distributions}\label{sec:family}
\subsection{Targets and reference data}
Let $H(\la)$ be a Hamiltonian with parameter vector $\la\in\Lambda\subset\mathbb R^{d_\lambda}$, where $\Lambda$ is the parameter domain considered and $d_\lambda$ is the number of Hamiltonian parameters. For a specified ground state $|\psi_0(\la)\rangle$, the target is
\begin{equation}
p_\la(\mathbf{x})=|\langle \mathbf{x}|\psi_0(\la)\rangle|^2,\qquad \mathbf{x}\in\{0,1\}^n.
\label{eq:target}
\end{equation}
This fixes the measurement basis. Matching $p_\la$ does not in general recover the phases of the wavefunction or establish quantum-state fidelity.

We use two main targets. The transverse-field Ising model (TFIM) is
\begin{equation}
H_{\rm I}=-J\sum_{\langle i,j\rangle}Z_iZ_j-g\sum_i X_i,
\label{eq:tfim}
\end{equation}
where the numerical benchmark uses a periodic $3\times3$ lattice. The second target is the open, cluster-type Haldane chain
\begingroup
\small
\begin{equation}
H_{\rm H}=-J\sum_{i=1}^{n-2}Z_iX_{i+1}Z_{i+2}-h_1\sum_{i=1}^n X_i-h_2\sum_{i=1}^{n-1}X_iX_{i+1},
\label{eq:haldane}
\end{equation}
\endgroup
with $\la=(h_1/J,h_2/J)$. This is the spin model studied in Ref.~\cCong, rather than the usual fermionic honeycomb Haldane model~\cite{haldane_model_1988}. We fit a separate model for each target family. Information is shared across its Hamiltonian parameters.

At each reference point $\la_j$, we obtain a ground state and draw a dataset $\mathcal D_j$ of computational-basis outcomes. Small systems use exact diagonalization. Larger chains use matrix product states (MPSs) and density-matrix renormalization group (DMRG)~\cite{schollwoeck_densitymatrix_2011,fishman_itensor_2021}. The latter supply approximate reference states whose accuracy depends on the solver and bond dimension. Training receives samples or selected correlators, whereas full probabilities are retained only for small-system evaluation. Appendix~\ref{app:targets} specifies the boundary conditions and how reference states are selected when the ground state is degenerate.

\subsection{A hypernetwork generates the circuit angles}
A hypernetwork is a neural network that supplies parameters of another model~\cite{ha_hypernetworks_2016,chauhan_brief_2024}. Related quantum approaches use neural networks to map Hamiltonian parameters to circuit angles~\cite{miao_neuralnetworkencoded_2024} or tensor-train hypernetworks to generate variational circuit parameters~\cite{qi_tensorhypervqc_2026}.
Following the ground-state tomography construction of Ref.~\cTonner, we map normalized Hamiltonian parameters to a scale and shift of a shared base vector,
\begin{equation}
\tht(\la)=[\mathbf1+\boldsymbol\gamma_\phiv(\la)]\odot\tht_0
+\boldsymbol\beta_\phiv(\la).
\label{eq:hyper}
\end{equation}
Here $\odot$ denotes componentwise multiplication, $(\mathbf a\odot\mathbf b)_j=a_jb_j$, and $\mathbf1$ is the vector of ones.
The scale-and-shift operation is also known as feature-wise linear modulation~\cPerez. We train the base circuit angles $\tht_0$ and hypernetwork weights $\phiv$ jointly, writing $\etav=(\tht_0,\phiv)$. At each Hamiltonian point, the hypernetwork receives the normalized coordinates of $\la$, such as the two coupling ratios specifying the Haldane Hamiltonian, and outputs the vectors $\boldsymbol\gamma_\phiv(\la)$ and $\boldsymbol\beta_\phiv(\la)$. Reference bitstrings and coefficients estimated from them are used to calculate the training loss, rather than being supplied as inputs to the hypernetwork. Each training update averages this loss over the available reference points and uses its gradients to update both $\tht_0$ and $\phiv$, as written explicitly in Appendix~\ref{app:joint-updates}. The reference data therefore guide the learned map from Hamiltonian parameters to circuit angles through these updates. The circuit architecture remains fixed as its angles change with $\la$.

Let $n_c$ be the total number of circuit qubits, including the $n$ measured qubits and $n_c-n$ unmeasured ancillas. Write $\mathbf{a}\in\{0,1\}^{n_c-n}$ for the ancilla outcomes. The model distribution is
\begin{equation}
q_{\etav,\la}(\mathbf{x})=\sum_{\mathbf{a}}|\langle \mathbf{x},\mathbf{a}|U_{\tht(\la)}|\psi_{\rm in}\rangle|^2.
\label{eq:born}
\end{equation}
We call the combined hypernetwork and QCBM model HyperQCBM.

We include ancillas to enlarge the family of distributions on the measured qubits. This is motivated by the universality of IQP Born machines with sufficiently many hidden qubits~\cite{kurkin_universality_2025}, although that result does not establish universality for our restricted circuits. Ancillas are marginalized in both training and evaluation. Once trained, the hypernetwork supplies circuit angles at a new parameter value without further optimization or reference data at that point. Quantum deployment consists of preparing $|\psi_{\rm in}\rangle$, applying the generated circuit and reading out the $n$ measured qubits. Before deployment, we can fine-tune the circuit using additional reference samples from the chosen Hamiltonian point. It starts from information learned across the diagram instead of fitting an independent circuit from its initial parameters.

\subsection{Circuit generators and conserved symmetries}\label{sec:circuits}
We use an ordered product of Pauli rotations,
\begin{equation}
U_\tht=e^{-i\theta_dP_d}\cdots e^{-i\theta_1P_1},
\label{eq:circuit}
\end{equation}
where each Hermitian Pauli string satisfies $P_j^2=I$. Lie algebras provide a way to select generators with specified conservation laws~\cite{kökcü_fixed_2022,wiersema_classification_2024}. This choice supplies an inductive bias by restricting the transformations available at every Hamiltonian point, independently of the fitted angles. If a Pauli symmetry commutes with every generator, the circuit preserves it throughout training and deployment. Its restriction on the output distribution also depends on the input state. We compare Haldane, Heisenberg and quadratic fermionic families with commuting $X$-string IQP circuits. The algebra determines which Pauli generators are allowed. The circuit model is defined by the generators we select and the order in which we apply their gates. A fixed gate sequence need not realize every unitary generated by the algebra.

For the Haldane family, permitted strings have an even number of $Y/Z$ factors on each of the two circuit sublattices. They commute with the corresponding products of $X$ operators. Heisenberg generators preserve global $X$ and $Z$ parity. Quadratic fermionic generators are Majorana bilinears, and IQP generators contain only $X$ and identity. Appendix~\ref{app:gates} gives compact algebra expressions, gate selection and parameter counts. Gate weight and pair distance are distinct restrictions.

Preparation matters as well as the generators. For the main Haldane and IQP studies we prepare the Greenberger--Horne--Zeilinger (GHZ) state $|\mathrm{GHZ}+\rangle=\frac{1}{\sqrt2}(|0\rangle^{\otimes n_c}+|1\rangle^{\otimes n_c})$.
If every gate commutes with global $X$, this preparation gives $q(\mathbf{x})=q(\bar{\mathbf{x}})$, including after ancilla marginalization. All odd-order $Z$ correlators on measured qubits then vanish. The use of an input symmetry sector follows the classical-training construction of Ref.~\cArmengol. A computational zero input would generally give a different family of distributions. GHZ preparation requires one Hadamard and $n_c-1$ controlled-NOT gates in a connected tree, with depth determined by connectivity. Other preparations used in the numerical comparisons are stated explicitly.

\section{Classical training with selected correlations}\label{sec:training}
\subsection{From likelihood to Fourier coefficients}
Our initial joint training of the hypernetwork and base circuit angles uses a KL-divergence objective. With reference samples and accessible model probabilities, this is implemented by minimizing the empirical negative log likelihood (NLL)
\begin{equation}
\mathcal L_{\rm NLL}(\etav)=-\frac1K\sum_{j=1}^K\frac1{N_j}
\sum_{\mathbf{x}\in\mathcal D_j}\log q_{\etav,\la_j}(\mathbf{x}).
\label{eq:nll}
\end{equation}
All reference points receive equal weight. The difference between this loss and the mean KL divergence from the empirical targets is their entropy, which is independent of $\etav$. This is a choice of training objective, not a requirement of the hypernetwork architecture. Its present implementation calculates the model probability of every possible outcome and is used as a small-system baseline.

To avoid calculating the probability of every possible outcome, write a distribution in its Walsh--Hadamard basis,
\begin{align}
c_S(p)&=\langle Z_S\rangle_p=\E_{\mathbf{x}\sim p}(-1)^{\sum_{i\in S}x_i},\label{eq:coeff}\\
p(\mathbf{x})&=2^{-n}\sum_{S\subseteq[n]}c_S(p)(-1)^{\sum_{i\in S}x_i}.
\label{eq:fourier}
\end{align}
Here $Z_S=\prod_{i\in S}Z_i$, with identity on the remaining qubits and ancillas. The support of a coefficient is the set $S$ of qubits on which $Z_S$ acts nontrivially. The order of a coefficient is the number of qubits in its support, $|S|$, and specifies how many measured bits enter the parity correlation. The empty support gives $c_\varnothing=1$. Target coefficients are estimated by sample averages of the same parity.

Selecting a set $\Omega$ lets us train on a subset of the $2^n$ Fourier coefficients. Several distributions can agree on $\Omega$~\cBorn. Shared pretraining uses information beyond what an MMD fit at a single Hamiltonian point can provide. The initial angles for local fine-tuning reflect how the selected correlations vary across related Hamiltonians. We test whether this also helps recover correlations beyond the training order. Classical shadows also use selected measurements to estimate properties~\cHuangShadow. Here the reference data come from a fixed measurement basis, and the estimator computes observables of the model circuit.

\subsection{Monte Carlo estimates for ordered Pauli circuits}\label{sec:mc}
We now turn to calculating the model coefficients without constructing the full output distribution. For IQP circuits built from commuting $X$ strings, coefficients can be estimated by averaging phase functions over random bitstrings~\cite{nest_simulating_2010,recio-armengol_train_2026a}. We extend this approach to ordered Pauli rotations so that classical coefficient-based training can also use noncommuting generator families, including those chosen to preserve the symmetries discussed in Sec.~\ref{sec:circuits}. We apply the gates in the order given in Eq.~\eqref{eq:circuit}, without assuming that they commute.

The idea is to write a coefficient $c_S=\langle\psi_{\rm in}|U_\tht^\dagger Z_S U_\tht|\psi_{\rm in}\rangle$ as the expectation of a quantity calculated from a random bitstring. Each bitstring selects two normalized vectors that we evolve within invariant subspaces of the circuit. The real part of their overlap gives one Monte Carlo sample, and averaging independent samples gives an unbiased estimate of $c_S$. These random bitstrings are inputs to a classical calculation, not samples from the model's output distribution.

The cost depends both on how many samples we need and on how much work each sample takes. The variance controls the number of samples needed for a specified additive accuracy. We show that the sampled overlap is bounded, giving a variance bound independent of system size. Evaluating each overlap can nevertheless be expensive because the vectors may have exponentially many entries. We therefore examine how the gate generators determine their size and which restrictions make the calculation efficient.

We first take $|\psi_{\rm in}\rangle=|0\rangle^{\otimes n_c}$, then extend the construction to the GHZ input used in our Haldane circuits. Appendices~\ref{app:mc-update} and~\ref{app:mc-cost} give the gate updates, estimator derivation and error bounds.

\paragraph*{Gate action in the $X$ basis.}
We work in the $X$ basis because $X$-string rotations only change phases there. We can then identify which basis states the remaining gates mix and how many amplitudes each Monte Carlo sample requires. The output is still measured in the computational basis.

We label each $X$-basis state of the complete circuit register by a bitstring $\mathbf{b}$ and write
\[
|\widetilde{\mathbf{b}}\rangle=H^{\otimes n_c}|\mathbf{b}\rangle,
\]
where $H$ is the single-qubit Hadamard gate. A bit $b_i=0$ denotes $|+\rangle_i$, while $b_i=1$ denotes $|-\rangle_i$. Each label covers all $n_c$ qubits, including the ancillas. Ancilla measurement outcomes are still denoted by $\mathbf{a}$.

To describe the gates, we use binary vectors in $\mathbb F_2^{n_c}$, where $\mathbb F_2=\{0,1\}$ with arithmetic modulo two. We write $\oplus$ for bitwise addition and evaluate binary dot products modulo two. Each Pauli word is specified by two such vectors, $\mathbf{x}_j$ and $\mathbf{z}_j$, through
\begin{equation}
P_j=i^{\nu_j}X^{\mathbf{x}_j}Z^{\mathbf{z}_j},\qquad
\nu_j=\sum_{i=1}^{n_c}(\mathbf{x}_j)_i(\mathbf{z}_j)_i.
\end{equation}
Here $X^{\mathbf{x}_j}=\bigotimes_i X_i^{(\mathbf{x}_j)_i}$ and $Z^{\mathbf{z}_j}=\bigotimes_i Z_i^{(\mathbf{z}_j)_i}$. A zero exponent contributes the identity in the corresponding tensor product. A $Y$ factor has both entries equal to one, with the required phase supplied by $i^{\nu_j}$. The sum defining $\nu_j$ is an ordinary integer sum, so it counts the $Y$ factors rather than their parity. Any overall minus sign in $P_j$ can be absorbed into the real rotation angle $\theta_j$.

In this basis, $X_i$ multiplies a basis state by $(-1)^{b_i}$, while $Z_i$ flips the bit $b_i$. Applying $P_j$ therefore changes the basis label from $\mathbf{b}$ to $\mathbf{b}\oplus\mathbf{z}_j$ and multiplies the state by a phase. Appendix~\ref{app:mc-update} gives that phase explicitly. The rotation $e^{-i\theta_jP_j}$ can consequently mix only the states labelled by $\mathbf{b}$ and $\mathbf{b}\oplus\mathbf{z}_j$. If $\mathbf{z}_j=0$, each basis state acquires only a phase.

Successive gates can change the initial bitstring only by adding their vectors $\mathbf{z}_j$ modulo two. We collect all possible binary sums of these vectors in
\begin{equation}
V=\operatorname{span}_{\mathbb F_2}\{\mathbf{z}_1,\ldots,\mathbf{z}_d\},\qquad r=\dim V.
\label{eq:rank}
\end{equation}
The dimension $r$ is the number of linearly independent vectors needed to generate these sums. Starting from $|\widetilde{\mathbf{b}}\rangle$, the circuit therefore remains within the span of basis states labelled by $\mathbf{b}+V=\{\mathbf{b}\oplus\mathbf{v}:\mathbf{v}\in V\}$. This set is called a coset and contains $2^r$ bitstrings. We simulate the evolution within one such subspace at a time, using a normalized vector with $2^r$ entries.

\paragraph*{Calculating a Monte Carlo sample.}
To calculate one Monte Carlo sample, we use a matrix $B\in\mathbb F_2^{n_c\times r}$ whose columns form a basis of $V$, chosen once for all samples. For a sampled bitstring $\mathbf{b}$, the vector entry indexed by $\mathbf{t}\in\mathbb F_2^r$ corresponds to the $X$-basis state specified by $\mathbf{b}\oplus B\mathbf{t}$. For zero input, every entry initially equals $2^{-r/2}$. Apply gates $1,\ldots,d$ in order to obtain the normalized vector $w_\tht^{(\mathbf{b})}(\mathbf{t})$.

Let $\mathbf{s}\in\mathbb F_2^{n_c}$ have a one at the position of each measured qubit in $S$ and zeros elsewhere, including the ancillas. The observable $Z_S$ maps the basis state specified by $\mathbf{b}$ to the one specified by $\mathbf{b}\oplus\mathbf{s}$. Evolve a second vector using bitstring $\mathbf{b}\oplus\mathbf{s}$, the same basis $B$ and the same angles. Pairing entries with the same $\mathbf{t}$ gives
\begin{align}
G_S(\mathbf{b};\tht)&=\operatorname{Re}\sum_{\mathbf{t}\in\mathbb F_2^r}
w_\tht^{(\mathbf{b}\oplus\mathbf{s})}(\mathbf{t})^*w_\tht^{(\mathbf{b})}(\mathbf{t}),\nonumber\\
\widehat c_S&=\frac1{N_{\rm MC}}\sum_{m=1}^{N_{\rm MC}}G_S(\mathbf{b}_m;\tht).
\label{eq:estimator}
\end{align}
Draw the $N_{\rm MC}$ bitstrings independently and uniformly from $\mathbb F_2^{n_c}$. Zero input has equal weight in every coset, and every coset contains the same number of bitstrings. Averaging $G_S$ therefore gives the desired coefficient. Since it is the real overlap of two normalized vectors, $|G_S|\leq1$, and for fixed angles
\begin{equation}
\begin{aligned}
\E\widehat c_S&=c_S,\\
\Var(\widehat c_S)&=\frac{\Var(G_S)}{N_{\rm MC}}\leq\frac1{N_{\rm MC}}.
\end{aligned}
\label{eq:variance}
\end{equation}
These are additive-error guarantees for the coefficient. They do not make a nonlinear loss of $\widehat c_S$ unbiased.

\begin{informalproposition}
For zero input and fixed circuit angles, Eq.~\eqref{eq:estimator} gives unbiased Fourier coefficient estimates, including for noncommuting circuits. With $N_{\rm MC}$ independent uniform draws, the root mean squared sampling error is at most $N_{\rm MC}^{-1/2}$, independently of the number of qubits. Each draw evolves two vectors with $2^r$ amplitudes, with a cost of $O(d2^r)$ for the $d$ gate updates after preprocessing. We restrict the gates to keep these vectors small, allowing us to estimate the correlations needed for training without storing the full quantum state. Proposition~\ref{prop:mc-estimator} in Appendix~\ref{app:mc} gives the full error and computational bounds.
\end{informalproposition}

\paragraph*{GHZ input.}
For GHZ input, the $X$-basis amplitudes are uniform on states specified by even-parity bitstrings, meaning bitstrings with an even number of ones, so $\mathbf1\cdot\mathbf{b}=0$, where $\mathbf1$ is the all-ones vector. If every gate vector $\mathbf{z}_j$ has even parity, the circuit preserves this support. For even $|S|$, sample bitstrings uniformly from this support and use the same estimator. For odd $|S|$, the coefficient is exactly zero. Corollary~\ref{cor:mc-ghz} proves that the same estimation bounds apply under this condition, which holds for our Haldane and IQP gates. Other input states or parity-changing gates require a separate treatment. For Haldane gates, restricting the active $Y/Z$ sites lowers $r$ but also introduces additional conserved $X$ products. The resulting expressivity change is tested in Fig.~\ref{fig:restrictions}.

\paragraph*{(i) Recovering the IQP estimator.}
For $P_j=X^{\mathbf{x}_j}$, every $\mathbf{z}_j$ is zero and each vector has a single entry. Writing
$\Phi_\tht(\mathbf{b})=\sum_j\theta_j(-1)^{\mathbf{x}_j\cdot \mathbf{b}}$ gives
$w_\tht^{(\mathbf{b})}=e^{-i\Phi_\tht(\mathbf{b})}$. The overlap in Eq.~\eqref{eq:estimator} becomes
\begin{equation}
\widehat c_S^{\rm IQP}=\frac1{N_{\rm MC}}\sum_{m=1}^{N_{\rm MC}}
\cos\!\left[2\sum_{j:\mathbf{x}_j\cdot \mathbf{s}=1}\theta_j(-1)^{\mathbf{x}_j\cdot \mathbf{b}_m}\right].
\label{eq:iqp-recovery}
\end{equation}
The general overlap estimator thus reduces to the IQP phase estimator. For zero input, sample $\mathbf{b}$ uniformly from all $n_c$-bit strings. For GHZ input and even-order coefficients, sample uniformly from bitstrings with an even number of ones.
Appendix~\ref{app:mc-iqp} derives the phase cancellation leading to Eq.~\eqref{eq:iqp-recovery}, and Appendix~\ref{app:mc-ghz} derives the GHZ sampling rule.

\paragraph*{(ii) Cost of the full Haldane gate set.}
We now determine $r$ for the Haldane gate set in which the allowed $Y/Z$ pairs connect all qubits within each circuit sublattice. This gives the number of amplitudes, $2^r$, that each vector must contain.

Number the circuit qubits from $1$ to $n_c$, including the ancillas at their assigned positions. The even and odd positions define the two sublattices, $E=\{j\in\{1,\ldots,n_c\}:j\text{ even}\}$ and $O=\{j\in\{1,\ldots,n_c\}:j\text{ odd}\}$. These sets refer to the complete circuit register, including ancillas. A gate with $Y$ or $Z$ on two sites in the same sublattice flips those two bits of the $X$-basis label. It therefore preserves the parity of that label on each sublattice. Any $X$ factors change phases without flipping bits.

When the permitted pairs connect each sublattice, possibly through intermediate sites, their binary sums generate every flip pattern with even parity on both sublattices. The span is therefore
\begin{equation}
V_{\rm H}=\{\mathbf{z}:\ \mathbf{z}\cdot\mathbf1_E=\mathbf{z}\cdot\mathbf1_O=0\},
\qquad r=n_c-2.
\end{equation}
Here $\mathbf1_E$ and $\mathbf1_O$ have ones on their respective sublattices and zeros elsewhere. Each nonempty sublattice imposes one independent parity constraint, leaving $n_c-2$ independent binary directions. Appendix~\ref{app:mc-haldane} proves this by constructing a basis within each connected sublattice. In the twelve-qubit circuit of Fig.~\ref{fig:haldane}, this gives $r=10$.

One Monte Carlo sample is still the real overlap of two normalized vectors, so the variance bound is unchanged. However, each vector now contains $2^{n_c-2}$ amplitudes. The cost of evaluating one sample with this method therefore grows exponentially with $n_c$. The restriction below reduces this cost by limiting the sites on which $Y/Z$ pairs can act.

\paragraph*{(iii) Restricting the active Haldane sites.}
We select nonempty qubit sets $A_E\subseteq E$ and $A_O\subseteq O$ within the two circuit sublattices (even and odd positions, respectively). We permit $Y/Z$ pair gates only within these selected sets, while retaining $X$-type gates throughout the circuit. We call the selected qubits active sites. Connecting each selected set through the permitted pair gates gives
\begin{equation}
r=|A_E|+|A_O|-2.
\label{eq:rank-examples}
\end{equation}
Two active sites per sublattice give $r=2$ and four entries per vector. Three give $r=4$ and sixteen entries. Choosing $O(\log n_c)$ active sites gives polynomial evaluation cost for a polynomial gate count. The gate restriction also changes the distributions the circuit can represent.
Appendix~\ref{app:mc-haldane} derives Eq.~\eqref{eq:rank-examples}, including the case of disconnected selected sets.

To see how these updates work, consider $X$-type gates together with a single $Z_uZ_v$ generator, where $u$ and $v$ belong to the same sublattice. The only possible flip changes both bits $u$ and $v$, so $r=1$ and each vector has two entries. For zero input, we initialize $w=(1,1)^{\mathsf T}/\sqrt2$. A rotation $e^{-i\alpha Z_uZ_v}$ then acts as
\begin{equation}
w\longleftarrow
\begin{pmatrix}\cos\alpha&-i\sin\alpha\\-i\sin\alpha&\cos\alpha\end{pmatrix}w.
\end{equation}
Here the basis matrix has one column, $B=\mathbf{z}=\mathbf{e}_u\oplus\mathbf{e}_v$, where $\mathbf{e}_u$ is one at site $u$ and zero elsewhere. An $X^{\mathbf{x}_j}$ rotation multiplies entry $t\in\{0,1\}$ by
$e^{-i\theta_j(-1)^{\mathbf{x}_j\cdot(\mathbf{b}\oplus t\mathbf{z})}}$.
After these updates in circuit order, one MC sample is
\begingroup
\small
\begin{equation}
G_S(\mathbf{b})=\operatorname{Re}\!\big[w^{(\mathbf{b}\oplus \mathbf{s})}(0)^*w^{(\mathbf{b})}(0)+w^{(\mathbf{b}\oplus \mathbf{s})}(1)^*w^{(\mathbf{b})}(1)\big].
\end{equation}
\endgroup
Thus the IQP scalar phase is replaced by a two-entry calculation in this noncommuting example. Appendix~\ref{app:mc-noncommuting} derives both update matrices and works through two- and three-gate examples using the overlap estimator and Pauli conjugation.

\subsection{CDF-KL on selected marginals}\label{sec:cdf}
CDF-KL extends the KL-based training formulation to selected correlations by comparing cumulative probabilities of small marginals. It provides an alternative to MMD when the full likelihood is unavailable. A set of coefficients can define a small marginal distribution without defining the full $n$-bit distribution. For $A\subset[n]$ with $|A|=\ell$, all coefficients supported in $A$ determine
\begin{equation}
p_A(\mathbf{z})=2^{-\ell}\sum_{S\subseteq A}c_S(p)(-1)^{\sum_{i\in S}z_i}.
\label{eq:marginal}
\end{equation}
We select a collection $\mathcal A$ of such subsets. We need every coefficient within each selected marginal, but need not use every $\ell$-site subset of the system.

For each marginal, order its bit strings by their fixed binary integer value. Let $F_{p,A}(t)=\sum_{\mathbf{z}\leq t}p_A(\mathbf{z})$, with $t=0,\ldots,2^\ell-2$. We compare these cumulative probabilities using
\begin{align}
d_{\rm B}(u,v)&:=u\log\frac uv+(1-u)\log\frac{1-u}{1-v},\\
\mathcal L_{\rm CDF}&:=\frac{\sum_{A\in\mathcal A}\sum_{t=0}^{2^\ell-2}
 d_{\rm B}(F_{p,A}(t),F_{q,A}(t))}
 {|\mathcal A|(2^\ell-1)}.
\label{eq:cdf}
\end{align}
Choosing the sizes and collection of marginals indirectly controls the number of distinct coefficients that must be estimated, accounting for overlap between marginals and coefficients fixed by symmetry. In our restricted-MMD comparisons, MMD uses this same fixed set, with weights determined by its bandwidth $\sigma$. In Gaussian-sampled MMD, $\sigma$ instead determines the probabilities of drawing coefficient supports. We can match the number of coefficient estimates in that comparison, but the sampled sets and orders need not coincide.

This is an average of Bernoulli KL divergences for threshold events. It changes which probabilities are compared, rather than approximating the full-distribution KL by an unbiased estimator. It is also distinct from the KL divergence of the marginal probability vectors. The bit ordering is part of its definition. We average this marginal loss across reference points to update the hypernetwork and base circuit angles, as in the likelihood formulation.

Noisy coefficients can reconstruct negative probabilities. Before forming each CDF, we project its reconstructed vector onto the probability simplex and mix in a small uniform component, applying the same procedure to the target marginal. This uniform mixing follows label smoothing~\cite{szegedy_rethinking_2015}, applied here to both marginals before calculating CDF-KL.
Here it keeps every compared cumulative probability strictly between zero and one, making the CDF-KL terms and their derivatives with respect to those probabilities finite. This allows gradient-based training without logarithmic singularities at zero-probability events. Smoothing changes the loss, and simplex projection remains only piecewise differentiable. An unbiased coefficient estimator therefore does not give an unbiased CDF-KL loss or gradient. We compare exact and MC coefficient training directly. Appendix~\ref{app:losses} states the projection, smoothing and effect of retaining only selected marginals.

\subsection{MMD and a combined objective}\label{sec:mmd}
We compare CDF-KL with MMD, which compares distributions through a positive kernel~\cGretton. Both losses can use the same selected coefficients, but compare different functions of them. For the Gaussian kernel on bit strings,
$k_\sigma(\mathbf{x},\mathbf{y})=\exp[-\|\mathbf{x}-\mathbf{y}\|_2^2/(2\sigma^2)]$, its squared value is
\begin{equation}
\MMD_\sigma^2(p,q)=\sum_{S\subseteq[n]}w_\sigma(S)[c_S(p)-c_S(q)]^2,
\label{eq:mmd}
\end{equation}
where $w_\sigma(S)=\rho^{|S|}(1-\rho)^{n-|S|}$ and
$\rho=(1-e^{-1/(2\sigma^2)})/2$. The bandwidth therefore sets a distribution over coefficient orders. This representation is used in classical IQP optimization~\cite{armengol_iqpopt_2026,recio-armengol_train_2026a}.

The kernel weights define the probabilities of choosing the supports. Larger $\sigma$ favors smaller supports and hence lower-order coefficients. Smaller $\sigma$ favors larger supports. In particular, the mean support size is $\mathbb E_{S\sim w_\sigma}|S|=n\rho$.

This bandwidth choice also affects trainability. For product circuits and suitable shallow circuits, choosing $\sigma^2=\Theta(n)$ avoids exponential loss concentration under the circuit and data assumptions of the corresponding trainability results~\cite{rudolph_trainability_2024}. Their kernel uses $2\sigma$ in the denominator, so their linear bandwidth scaling corresponds to $\sigma=\Theta(\sqrt n)$ in our convention. This is a sufficient regime under those assumptions, rather than a universal requirement for successful training.

Two independent MC estimates give an unbiased product
$(\widehat c_S^{(1)}-c_S(p))(\widehat c_S^{(2)}-c_S(p))$
for a fixed empirical target coefficient. Squaring one noisy estimate would add its variance. In the matched comparisons, the two factors split the stated MC budget equally.

We also train with the combined loss $\mathcal L_{\rm CDF}+\alpha_{\rm MMD}\mathcal L_{\rm MMD}$, where $\alpha_{\rm MMD}\geq0$ sets the weight of the MMD term relative to CDF-KL. MMD weights squared coefficient differences through its Gaussian kernel, while CDF-KL compares cumulative probabilities within separately chosen, equally weighted marginals. Appendix~\ref{app:mmd-training} gives the kernel expansion and MC estimators, and Appendix~\ref{app:additional} reports the additional fixed-grid comparisons.

\section{Distribution learning and phase signatures}\label{sec:results}
We first test the distributions represented by the classically trained circuits where exact small-system evaluation is available. The comparisons address sampling accuracy, the effect of the circuit architecture, and the recovery of changes across the phase diagram. The small system sizes $n$ considered in this section allow us to evaluate these quantities by summing over all $2^n$ measurement bitstrings.
\subsection{Evaluation metrics and response functions}\label{sec:metrics}
For the small systems, we calculate distribution infidelity and total variation distance (TVD)~\cTao from all outcomes,
\begin{align}
\mathcal I(p,q)&=1-\left(\sum_{\mathbf{x}}\sqrt{p(\mathbf{x})q(\mathbf{x})}\right)^2,\\
\operatorname{TVD}(p,q)&=\frac12\sum_{\mathbf{x}}|p(\mathbf{x})-q(\mathbf{x})|.
\label{eq:metrics}
\end{align}
They obey~\cFuchs
\begin{equation}
1-\sqrt{1-\mathcal I}\leq\operatorname{TVD}\leq\sqrt{\mathcal I}.
\label{eq:bounds}
\end{equation}
We use infidelity as our main measure of distribution error because its local variation defines the distribution susceptibility used below.

For a differentiable distribution family we define
\begin{equation}
\chi_{\mu\nu}(q_\la)=\frac14\sum_{\mathbf{x}:q_\la(\mathbf{x})>0}
\frac{\partial_\mu q_\la(\mathbf{x})\partial_\nu q_\la(\mathbf{x})}{q_\la(\mathbf{x})}.
\label{eq:chi}
\end{equation}
It is one quarter of the classical Fisher information matrix. Locally,
$\mathcal I(q_\la,q_{\la+d\la})=\sum_{\mu\nu}\chi_{\mu\nu}d\lambda_\mu d\lambda_\nu+o(\|d\la\|^2)$ under regularity of the support (Appendix~\ref{app:susceptibility}). Fidelity responses have long been used to study phase transitions~\cite{zanardi_ground_2006,gu_fidelity_2010}. Here the response concerns a specified measurement distribution, not the full quantum-state fidelity. Derivatives of the learned model hold its trained weights fixed.

\begin{figure*}[!t]
\includegraphics[width=\textwidth]{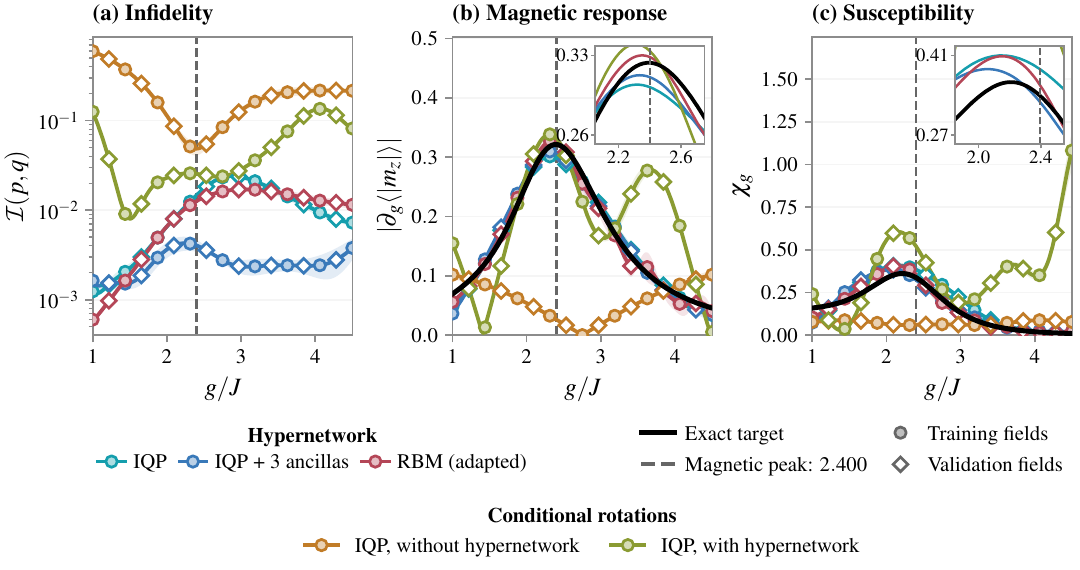}
\caption{\textbf{Distribution learning across the field of the periodic $3\times3$ TFIM.} (a) Distribution infidelity $\mathcal I(p,q)$, (b) magnetic response and (c) distribution susceptibility. Models include hypernetwork IQP with and without three unmeasured ancillas, the adapted RBM, and IQP with conditional rotations with and without a hypernetwork. Black curves show exact-target responses, and the dashed line marks the finite-size magnetic-response peak. Circles and open diamonds mark training and validation fields. Curves and bands show means and one sample standard deviation across three initializations with shared reference data. Appendix~\ref{app:training} specifies the IQP generators, GHZ preparation, $R_y$ encoding and RBM conditioning.}
\label{fig:tfim}
\end{figure*}

Phase transitions are commonly investigated through order parameters, their derivatives and fluctuations, and correlation functions. Magnetic susceptibility and heat capacity are familiar response indicators. Fidelity susceptibility and Fisher information characterize changes in states or measurement distributions without specifying an order parameter~\cite{gu_fidelity_2010,prokopenko_relating_2011}. Data-driven approaches also construct indicators from sampled configurations, using principal-component analysis~\cite{wang_discovering_2016}, supervised classification and learning by confusion~\cite{carrasquilla_machine_2017,vannieuwenburg_learning_2017a}, or derivatives of inferred physical parameters~\cite{schäfer_vector_2019}. Changes in mean input features provide another interpretable option~\cite{arnold_interpretable_2021}. Analytical studies express optimal classification and regression indicators in terms of the underlying distributions~\cite{arnold_replacing_2022}, extend this approach to generative classifiers~\cite{arnold_mapping_2024}, and bound several normalized indicators by the square root of the classical Fisher information~\cite{arnold_machine_2023}. ClassiFIM further estimates the Fisher information from sampled configurations through an auxiliary classification task~\cite{kasatkin_classifim_2024}. These connections place the distribution susceptibility in Eq.~\eqref{eq:chi} within a broader set of phase-transition indicators.

A phase indicator calculated from bitstrings can probe correlations beyond those available to the classical training procedure~\cBorn. If the indicator and its parameter response depend only on classically accessible correlators, they can be evaluated directly. For indicators estimated from samples, a comparison of classical and quantum approaches would need to account for the cost of generating those samples as well as training the model and evaluating the indicator.

\subsection{A one-parameter TFIM benchmark}\label{sec:tfim-results}
Figure~\ref{fig:tfim} tests whether classical training across the field of the nine-site periodic TFIM supplies accurate sampling circuits between reference points. Nine training fields span $g/J\in[1,4.5]$, with 5,000 samples per field and eight separate validation fields. We train for 1,500 likelihood updates and select checkpoints by empirical validation NLL.

Single-site base angles are drawn uniformly from $[0.2,0.35]$ and pair angles from a zero-mean Gaussian with standard deviation $0.03$. The hypernetwork initially introduces only small angle modulations. Initialization and circuit architecture both influence the trainability of IQP generative models~\cite{lerch_iqp_2026,shen_characterizing_2026}. Appendix~\ref{app:initialization} gives the likelihood-training settings.

We compare hypernetwork IQP with a restricted Boltzmann machine (RBM) adapted to the hypernetwork setting, extending earlier RBM--Born-machine comparisons~\cCoyleFinance. Over the 144 test fields and three initializations, their mean distribution infidelities are similar, at $0.0115$ and $0.0106$, respectively. Three unmeasured ancillas reduce the IQP infidelity to $0.00263$, a reduction by a factor of approximately $4$ relative to IQP without ancillas and the RBM. The circuit angle count increases from 45 to 78. Appendix~\ref{app:training} specifies the circuit preparations.

\begin{figure*}[!p]
\includegraphics[width=\textwidth]{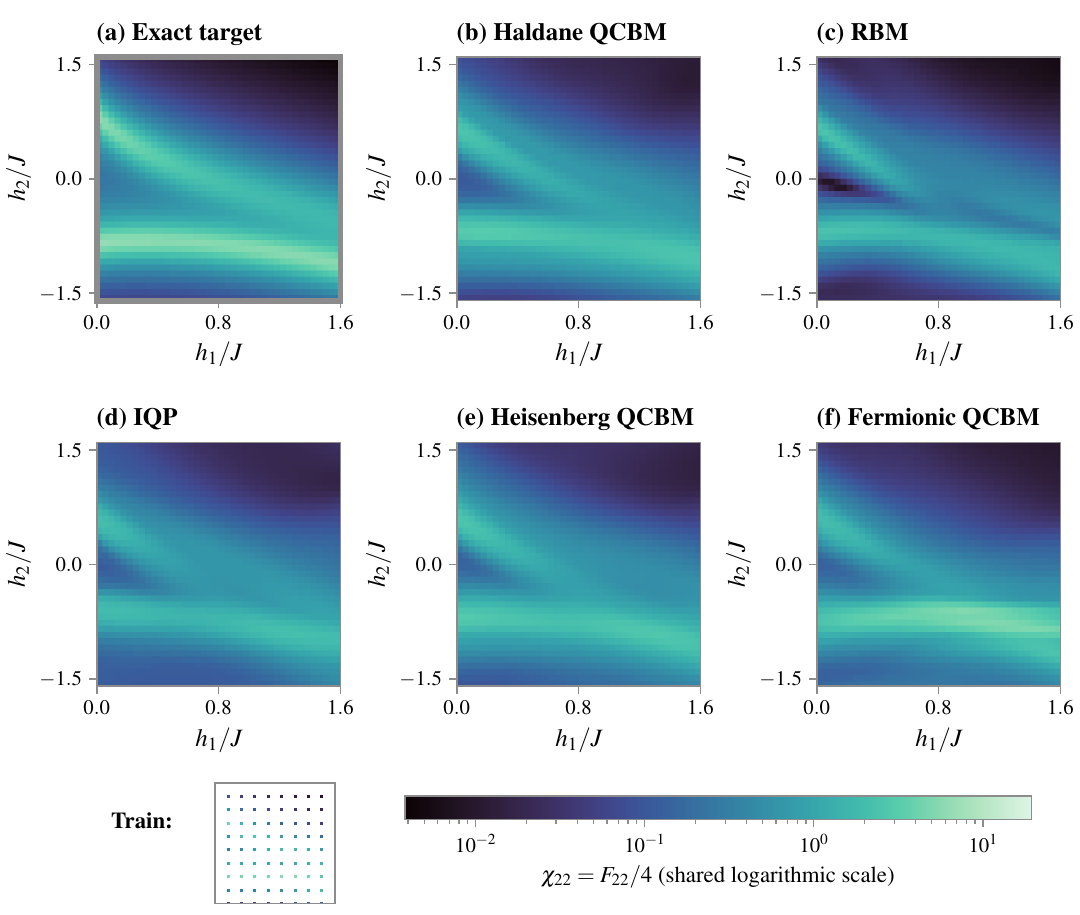}
\caption{\textbf{Learned distribution susceptibility for the nine-site open Haldane chain.} Colours show $\chi_{22}$ from Eq.~\eqref{eq:chi}, the response to $h_2/J$ at fixed $h_1/J$. Panels compare (a) the exact target with (b) the Haldane QCBM, (c) RBM, (d) IQP, (e) Heisenberg QCBM and (f) Fermionic QCBM. All panels use a common colour scale and evaluation grid. The inset shows the $10\times10$ training grid, coloured by target susceptibility. Each model uses 5,000 samples per training point and one initialization. Responses are calculated from all outcome probabilities after training.}
\label{fig:haldane}
\par\medskip
\begin{minipage}[t]{0.48\textwidth}
\vspace{0pt}
\makeatletter\def\@captype{table}\makeatother
\centering\begin{tabular}{lrr}\toprule
 & QCBMs & RBM\\\midrule
Generator parameters &298&299\\
Hypernetwork parameters &9,581&9,714\\
Total &9,879&10,013\\\bottomrule
\end{tabular}
\caption{\textbf{Parameter counts for Fig.~\ref{fig:haldane}, including all circuit angle parameters and hypernetwork weights.} Circuit families share the displayed counts. The RBM generator count includes its interaction matrix and biases.}
\label{tab:parameters}
\end{minipage}
\hfill
\begin{minipage}[t]{0.48\textwidth}
\vspace{0pt}
\makeatletter\def\@captype{table}\makeatother
\centering
\begin{tabular}{lrr}
\toprule
Model & Mean $\mathcal{I}(p,q)$ & Mean TVD\\
\midrule
Haldane QCBM & 0.00670 & 0.0436\\
RBM & 0.0447 & 0.124\\
IQP & 0.0182 & 0.0682\\
Heisenberg QCBM & 0.0126 & 0.0624\\
Fermionic QCBM & 0.0329 & 0.0929\\
\bottomrule
\end{tabular}
\caption{\textbf{Mean distribution errors over the held-out Haldane grid in Fig.~\ref{fig:haldane}.} Values refer to one fit per model.}
\label{tab:errors}
\end{minipage}
\end{figure*}

The hypernetwork learns field-dependent IQP angles through a nonlinear map. We also test conditional input rotations, as in Born-machine extensions such as~\cKiss, using a fixed affine field encoding through $R_y$ on each qubit. The IQP angles are either shared across fields or supplied by a hypernetwork, giving mean test infidelities of $0.206$ and $0.0514$, respectively. The hypernetwork therefore improves the rotation model, but neither variant improves on hypernetwork IQP without input rotations. This supports learning the Hamiltonian-to-circuit parameter map, which need not be linear. The comparison does not isolate nonlinearity, since the input rotations also change the circuit and can break its global-flip symmetry.

To test whether the model reproduces changes between training fields, we use two types of indicators discussed in Sec.~\ref{sec:metrics}, an order-parameter response and the distribution susceptibility. The magnetic response $|\partial_{g/J}\E_q|m_z||$, where $m_z=n^{-1}\sum_i(-1)^{x_i}$, tracks changes in magnetization as the transverse field suppresses magnetic order. Taking $|m_z|$ avoids cancellation between opposite magnetizations. The distribution susceptibility probes changes in all outcome probabilities, including those that the mean magnetization magnitude may miss. Low infidelity alone does not ensure accurate responses, so we compare both curves and their peaks with those of the exact nine-site system.

\begin{figure}[!t]
\includegraphics[width=\columnwidth]{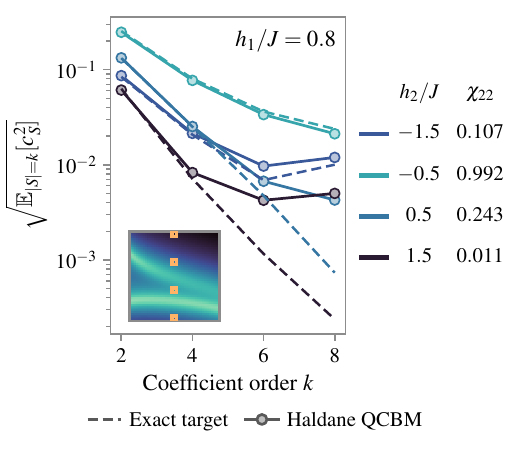}
\caption{\textbf{Fourier RMS spectra at four untrained points.} Solid curves with circular markers show the Haldane QCBM from Fig.~\ref{fig:haldane}(b). Dashed curves show the exact targets. Both spectra show the RMS defined in Eq.~\eqref{eq:rms}, using exact probabilities.}
\label{fig:fourier}
\end{figure}

\subsection{Learning the Haldane phase diagram}\label{sec:haldane-results}
Near the Haldane symmetry-protected topological phase boundary, the correlations used to identify the phase become weak, so more measurements are needed to distinguish them from sampling noise~\cCong. Finite system sizes also smooth the transition~\cite{herrmann_realizing_2022}.

Figure~\ref{fig:haldane} extends the task to two Hamiltonian parameters and compares the inductive biases of different generator families under the same reference data. The nine-site target uses $h_1/J\in[0,1.6]$ and $h_2/J\in[-1.6,1.6]$. A fixed $10\times10$ grid supplies 100 reference datasets, with 5,000 samples each. We use 24 separate validation points and evaluate on a held-out $40\times40$ grid. This comparison uses one initialization and 1,500 likelihood updates per model.

We use the distribution susceptibility $\chi_{22}$ of Eq.~\eqref{eq:chi}, evaluated with respect to $h_2/J$ at fixed $h_1/J$. Comparing the learned and exact-target maps tests whether the model reproduces the finite-chain phase-transition signatures.

The circuits act on nine measured qubits and three interleaved ancillas. Each has 298 circuit angle parameters, with generators and repetitions chosen as in Appendix~\ref{app:gates}. We count one angle per parametrized gate, including repeated generators and gates whose angles do not affect the distribution of measured outcomes. Table~\ref{tab:parameters} separates these angles from hypernetwork weights. Circuits with the same number of angle parameters can represent different sets of distributions. The RBM has approximately matched generator and hypernetwork budgets.

The Haldane QCBM reproduces the two prominent response bands of the exact target in Fig.~\ref{fig:haldane}(a,b), although their positions and widths do not agree everywhere. The other models also recover broad response features, with differences in their location and contrast. These maps assess the parameter dependence of the learned distributions, beyond their pointwise accuracy.

Table~\ref{tab:errors} shows that the Haldane QCBM gives the lowest mean infidelity and TVD, $0.00670$ and $0.0436$, respectively. The Heisenberg QCBM is next in both metrics, with $0.0126$ and $0.0624$, followed by IQP, the Fermionic QCBM and the RBM. Thus equal circuit angle counts do not give equal accuracy in this comparison. These observations concern one fit per model with the stated budgets. Additional target Hamiltonians are considered in Appendix~\ref{app:additional}.

\subsection{Fourier coefficients by correlation order}\label{sec:spectrum}
Distribution error averages over all outcomes. The Fourier spectrum gives a different view by separating correlations by order. At a fixed parameter point, we calculate the root mean square (RMS) coefficient magnitude
\begin{equation}
\sqrt{\E_{|S|=k}[c_S^2]}=
\left[\binom nk^{-1}\sum_{|S|=k}c_S^2\right]^{1/2}.
\label{eq:rms}
\end{equation}
The average is uniform over subsets of size $k$. It is not an average over training repetitions. Here and in Fig.~\ref{fig:transfer}, we calculate the model and target probabilities of every possible outcome to obtain the evaluation coefficients. The plotted RMS and squared errors therefore contain no MC evaluation noise, even when the model was trained with MC estimates.

Figure~\ref{fig:fourier} compares the Haldane model with its target at four points on $h_1/J=0.8$. None belongs to the fixed training grid. This is a test of interpolation after likelihood training. The selected-coefficient training comparison follows in Sec.~\ref{sec:sharing}. Only even orders are shown because the preparation and circuit enforce bit-complement symmetry. The spectrum reveals where the learned correlation scale departs from the target, including at high orders with small exact coefficients. Different sets of coefficients can have the same RMS.

The closest relative agreement across the displayed orders occurs at $h_2/J=-0.5$, the point with the highest target susceptibility in Fig.~\ref{fig:fourier} ($\chi_{22}\simeq0.992$). Here the learned RMS follows the target through order eight. At the lower-susceptibility points $h_2/J=0.5$ and $1.5$, the model overestimates the RMS at orders six and eight, despite close agreement at lower orders.

\subsection{Locating rapid changes in the learned family}\label{sec:boundary}
Spin correlators can provide interpretable signatures of phase structure. For example, TetrisCNN detects phases from measurement snapshots and expresses order parameters in terms of spin correlators~\cite{cybiński_speak_2024}. Here, we use the response of the learned distribution to locate candidate sampling points near phase boundaries, without identifying an order parameter.

To choose where to sample near a transition, we first need to locate rapid changes in the learned distributions. Learning a family gives a differentiable response throughout the diagram. We identify candidate boundaries by locating interior local maxima of $\chi_{22}$ along vertical slices. Adjacent maxima are linked only when their separation is within the grid tolerance specified in Appendix~\ref{app:peaks}, to avoid joining distant peaks into an artificial boundary. This procedure uses no phase labels and does not force the model peaks onto a reference curve.

\begin{figure}[!htbp]
\includegraphics[width=\columnwidth]{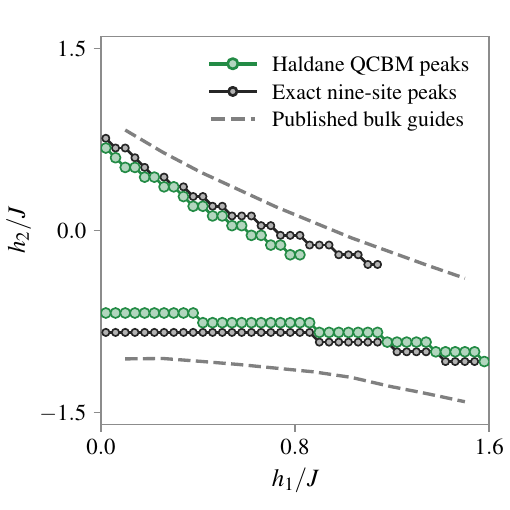}
\caption{\textbf{Boundary detection from the Haldane QCBM fit in Fig.~\ref{fig:haldane}(b).} Learned susceptibility peaks are compared with exact finite-size peaks and the bulk guides.}
\label{fig:boundary}
\vspace{-18pt}
\end{figure}

Figure~\ref{fig:boundary} applies the procedure to the Haldane model already fitted in Fig.~\ref{fig:haldane}(b). The reconstructed boundaries lie close to the exact nine-site susceptibility peaks along both branches. Published bulk boundaries in Ref.~\cCong supply a separate physical guide. Those curves were obtained numerically from the curvature of the infinite-chain energy density,
\begin{equation}
e_\infty(u,v)=\lim_{n\to\infty}\frac{E_0^{(n)}(u,v)}n,
\label{eq:bulk}
\end{equation}
where $u=h_1/J$ and $v=h_2/J$. The boundaries are identified by singular or peaked $|\partial_v^2e_\infty|$ along a slice. A finite-size probability-response peak and a bulk energy-curvature boundary are different reference quantities.

\begin{figure*}[!t]
\includegraphics[width=\textwidth]{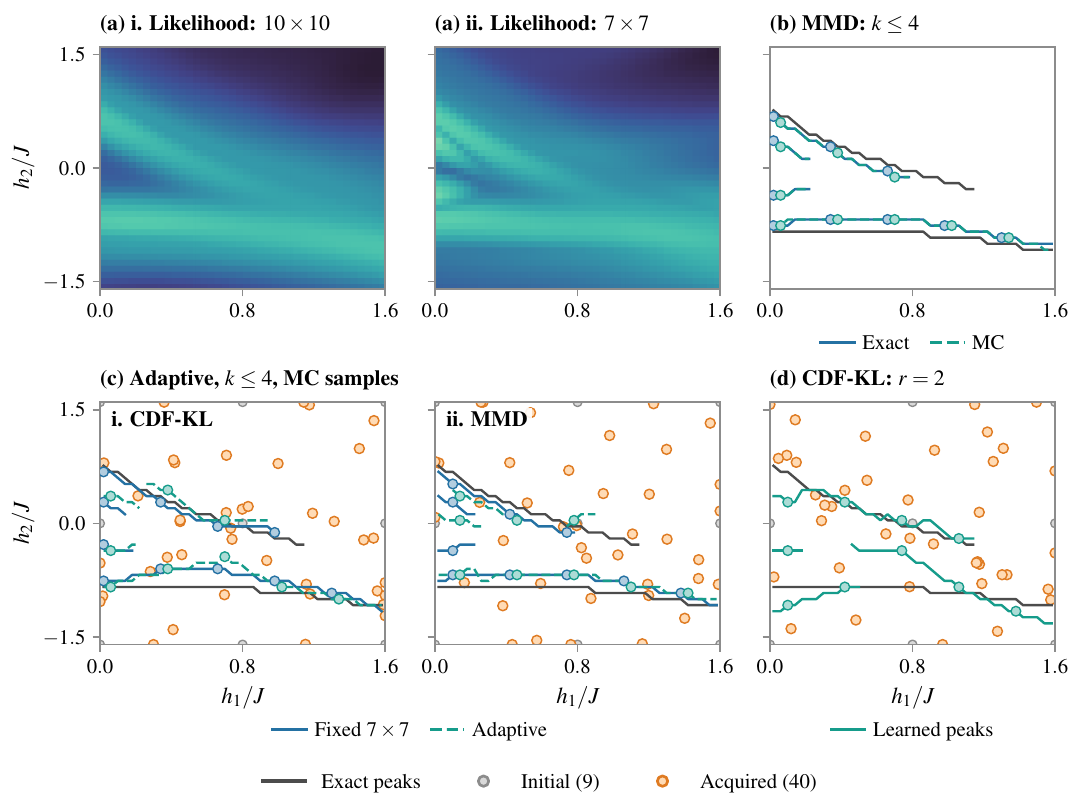}
\caption{\textbf{Effects of restricted training for the nine-site Haldane target.} All models are Haldane QCBMs. (a.i,ii) Susceptibility maps after likelihood training on $10\times10$ and $7\times7$ grids, with the colour scale of Fig.~\ref{fig:haldane}. (b) Fixed-grid MMD with exact or MC coefficients through order four. (c.i,ii) Fixed and adaptive 49-point fits using CDF-KL and MMD, respectively, with the same order limit and MC budget. (d) Adaptive CDF-KL with restricted gates giving $r=2$. Turquoise curves mark adaptive learned peaks, gray curves exact finite-size peaks, gray points the nine initial references, and orange points the 40 acquired references. Panels retain 298 circuit angle parameters and 9,581 hypernetwork weights. All plotted responses are evaluated from full model probabilities.}
\label{fig:restrictions}
\end{figure*}

\subsection{Effects of restricting classical training}\label{sec:restrictions}
We now test whether the response can still be recovered when classical training has fewer reference states and access to only selected correlations. Figure~\ref{fig:restrictions} first reduces the likelihood-training grid from $10\times10$ to $7\times7$. This retains full model probabilities and isolates the smaller reference grid within that comparison. Mean held-out infidelity increases from $0.00670$ to $0.0175$.

We then train on coefficients through order four. CDF-KL uses all 126 four-bit marginals. Their union contains 255 nonconstant coefficients. Of these, 162 even-order model coefficients require evaluation, while odd coefficients vanish by symmetry. The order-four MMD comparison uses the same coefficient information with its kernel weights. MC versions use 64 draws per nontrivial coefficient per update. All susceptibilities in Fig.~\ref{fig:restrictions} are evaluated from exact model probabilities, including for models trained with MC estimates.

The fixed-grid CDF-KL infidelity is $0.0743$ with exact coefficients and $0.0751$ with MC estimates. The corresponding MMD values are $0.0497$ and $0.0496$. Under these settings, changing the coefficient estimator has a smaller effect on mean infidelity than replacing full likelihood by a restricted loss. The final panel further restricts the allowed gates to give $r=2$, so each simulated vector has four amplitudes. Its loss still uses order-four coefficients, but the gate restriction changes the set of distributions the circuit can represent.

We describe how the adaptive fits select reference points in Sec.~\ref{sec:acquisition} and apply the procedure to larger systems in Sec.~\ref{sec:large-systems}. Appendix~\ref{app:additional} gives the full error table and compares gate restrictions.

\section{Adaptive reference acquisition}\label{sec:acquisition}
\subsection{Choosing where to calculate the next ground state}
Each new reference point requires a separate ground-state calculation. For large systems, this cost limits how many distributions can contribute to classical training. Our purpose is to select a sparse set of useful references, with particular interest in regions near phase transitions. A dense grid can spend most of its cost in regions where the distribution changes slowly. Earlier machine-learning studies locate phase changes from labelled or unlabelled data~\cite{carrasquilla_machine_2017,vannieuwenburg_learning_2017a,wang_discovering_2016}. Latent representations can expose phase structure~\cWetzel, and symbolic methods can connect these representations to physical observables~\cite{schoulepnikoff_discovering_2026}. Autonomous acquisition can use latent representations to guide differentiable ground-state calculations~\cZhou. We adapt the latent-distance search to use derivatives of our learned correlators, avoiding differentiation through the reference solver.

Let $y_p(\la)$ contain a fixed set of target coefficients at acquired points, standardized using those data. We fit an autoencoder with encoder $e_\psi$ and decoder $d_\psi$ by reconstruction error. We encode the coefficients at each acquired reference point and take their average in latent space,
\begin{equation}
\mu_t=\frac1{|\mathcal D_t|}\sum_{\la_j\in\mathcal D_t}e_\psi(y_p(\la_j)).
\end{equation}
For a candidate point, the trained circuit supplies $y_q(\la)$. We define
\begin{equation}
A_t(\la)=-\frac{\sqrt{\|e_\psi(y_q(\la))-\mu_t\|_2^2+\delta^2}}{\sqrt{d_z}},
\label{eq:acquisition}
\end{equation}
where $d_z$ is the latent dimension and $\delta>0$ regularizes the norm. After rescaling each Hamiltonian parameter to the interval $[0,1]$, we update the resulting vector $u$ according to
\begin{equation}
u_{s+1}=\Pi_{[0,1]^{d_\lambda}}\left[u_s-\alpha
\frac{\nabla_u A_t(u_s)}{\|\nabla_u A_t(u_s)\|_2+\epsilon}\right].
\label{eq:acqstep}
\end{equation}
Here, $\Pi_{[0,1]^{d_\lambda}}$ projects onto the normalized parameter domain by clipping each coordinate to $[0,1]$.
The search moves the encoded model prediction away from this average. We also sample less explored regions and require new points to be sufficiently far from existing ones, so the search does not concentrate on a single region. We then calculate a reference state at the accepted point, add its data, and refit the shared model. Appendix~\ref{app:acquisition} gives the search and training settings.

In the nine-site fits, some acquired points cluster along branches of the learned response in Fig.~\ref{fig:restrictions}. With 49 references, the adaptive CDF-KL and MMD fits have mean infidelities of $0.0819$ and $0.0655$ over the full grid, compared with $0.0751$ and $0.0496$ for their fixed-grid counterparts. These averages describe accuracy across the diagram. The selected coordinates show where the search concentrates the reference calculations.

\begin{figure*}[!t]
\centering
\includegraphics[width=\textwidth]{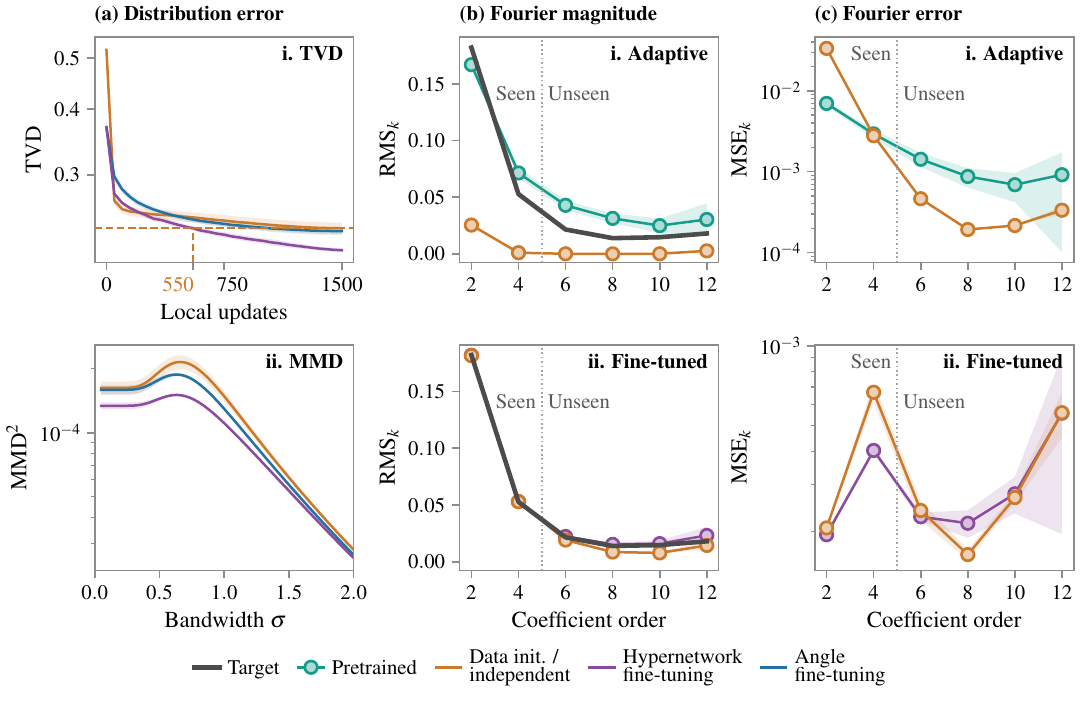}
\caption{\textbf{Shared training and local refinement for the twelve-site Haldane model with three ancillas.} (a.i) TVD during independent training, angle-only fine-tuning and hypernetwork fine-tuning. The guides mark when hypernetwork fine-tuning reaches the independent fits' final mean TVD. (a.ii) Final squared MMD across Gaussian bandwidths, evaluated using all coefficient orders. (b) Fourier RMS and (c) coefficient MSE, defined in Eqs.~\eqref{eq:transfer-rms}--\eqref{eq:transfer}. Upper panels compare adaptive pretraining with data-informed initialization, and lower panels compare hypernetwork fine-tuning with trained independent models. The separator distinguishes orders used in training ($k\leq4$, Seen) from higher orders (Unseen). Local fits use the same samples and 1,500 CDF-KL plus MMD updates, excluding pretraining. Curves show means and bands one sample standard deviation across four initializations, with 40 common boundary points averaged within each initialization. Evaluation uses exact model and target probabilities.}
\label{fig:transfer}
\end{figure*}

\section{Sharing information and refining boundary samplers}\label{sec:sharing}
\subsection{A common set of boundary targets}
The low-order loss does not directly compare higher-order coefficients with their targets. These coefficients still depend on the shared parameters, generator choice and preparation. We ask whether fitting related distributions across the diagram helps recover them at a selected boundary point, compared with fitting that point independently.

Figure~\ref{fig:transfer} uses a twelve-site Haldane target, three ancillas and an $r=4$ circuit with 316 circuit angle parameters. Fixed-grid and adaptive hypernetworks each train on 49 reference points. The loss uses all 495 four-bit marginals, requiring 561 nonzero even-order coefficients estimated with 64 MC draws each. No coefficient above order four enters this loss.

We freeze 40 coordinates on the adaptive model's learned susceptibility boundary. At every one of these same coordinates we compute a new target and evaluate both pretrained hypernetworks. We also train an independent QCBM directly on its local data, with the same circuit, local samples and 1,500 updates as fine-tuning. Pretraining uses CDF-KL, while all local fits use CDF-KL plus MMD restricted to orders through four. The pretrained hypernetworks use data from other Hamiltonian points, while the independent models are trained at the selected boundary points.

For each model and initialization, we calculate the RMS coefficient magnitude and mean squared error (MSE) of the coefficients,
\begin{align}
\mathrm{RMS}_k&=\sqrt{\E_{\la\in\mathcal B}\E_{|S|=k}[c_S(q_\la)^2]},\label{eq:transfer-rms}\\
\mathrm{MSE}_k&=\E_{\la\in\mathcal B}\E_{|S|=k}
[(c_S(q_\la)-c_S(p_\la))^2].
\label{eq:transfer}
\end{align}
Both averages are uniform. The squared error is calculated before averaging, so it retains coefficient-wise differences that an RMS spectrum can hide. The small-system evaluation uses every coefficient of each displayed order, including those absent from training. Across initializations, we keep the training and validation data, acquisition points and order, MC draws and boundary points fixed. Figure~\ref{fig:transfer-learning} in Appendix~\ref{app:finetuning} shows the susceptibility maps, learned boundary points and pretraining loss curves for the four initializations.

Before local fitting, the pretrained model already captures much of the target's low-order correlation strength in Fig.~\ref{fig:transfer}(b.i). At orders two and four, its gap from the target RMS is approximately ten and three times smaller, respectively, than that of the data-informed initialization. The initialization gives nearly zero RMS above order two, while pretraining produces nonzero higher-order correlations.

We evaluate coefficient errors above order four to test how well the learned circuit reproduces correlations outside the training loss.
Sample-based coverage tests measure a complementary quantity, the fraction of previously unseen valid outcomes reached within a given sampling budget~\cite{gili_generalization_2023,gili_quantum_2023,raj_train_2026}.

\begin{figure*}[p]
\centering
\includegraphics[width=0.99\textwidth]{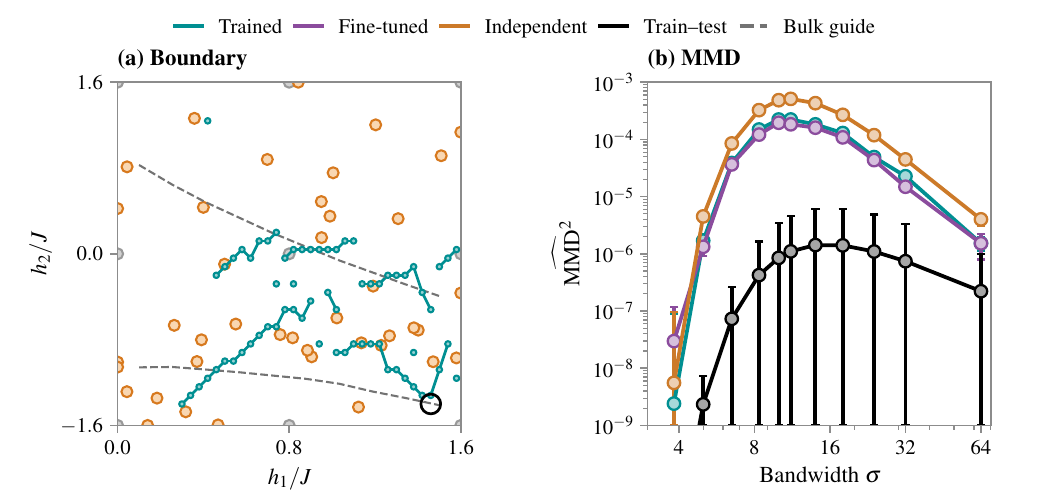}
\caption[Training and refinement for the 1,024-site chain]{\textbf{Shared training and local refinement for an open 1,024-site Haldane chain.} (a) Peaks in the learned marginal response, with nine initial references (gray), 41 acquired references (orange) and the published bulk boundaries (dashed)~\cCong. The ring marks the local fitting coordinate. (b) Bias-corrected squared MMD against one million held-out samples from the reference MPS (bond cap 128). The black curve compares the 20,000 local training samples with the held-out samples. Bars show standard errors across 40 groups.}
\label{fig:large}
\par\smallskip
\includegraphics[width=0.99\textwidth]{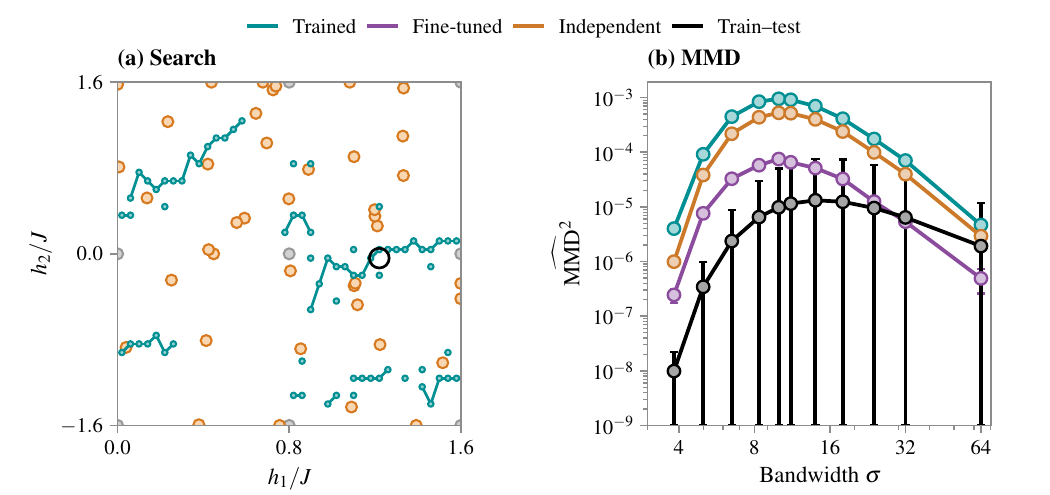}
\caption[Training and refinement for an open rectangular spin lattice]{\textbf{Shared training and local refinement for an open 4 by 256 spin lattice.} (a) Peaks in the learned marginal response, with nine initial references (gray) and 41 acquired references (orange). The ring marks the local fitting coordinate. (b) Bias-corrected squared MMD against one million held-out samples from the reference MPS (bond cap 128). The black curve compares the 20,000 local training samples with the held-out samples. Bars show standard errors across 20 groups.}
\label{fig:large-2d}
\end{figure*}

\subsection{Local fine-tuning from the learned family}
Interpolation supplies a circuit without a new local fit. When reference data at the selected point are available, fine-tuning provides a second route to preparing it for quantum deployment. For each boundary coordinate, we make a fresh copy of the pretrained hypernetwork and base angles and optimize both on the local data. All copies within a variant start from the same pretrained parameters and fresh optimizer moments. No parameters pass between boundary points. Fine-tuning and independent training use 1,500 local updates with CDF-KL plus MMD and the same local samples. The learning rate decays linearly from 0.003 to 0.00003. Angle-only fine-tuning starts from the pretrained circuit's 316 output angles and optimizes them directly. Hypernetwork fine-tuning instead updates the hypernetwork and shared base angles, a total of 10,473 parameters. We perform this separately for the fixed-grid and adaptive models. Appendix~\ref{app:finetuning} gives the numerical settings.

Figure~\ref{fig:transfer}(a.i) shows that adaptive hypernetwork fine-tuning improves on the final TVD of independent training. Across the four initializations, its final mean TVD is 0.216, compared with 0.237 for independent training and 0.234 for angle-only fine-tuning. It matches the independent fits' final mean TVD after approximately 550 local updates, compared with their 1,500 updates. Its final mean TVD is lower than independent training in each initialization.

The improvement also appears in Fig.~\ref{fig:transfer}(a.ii). Hypernetwork fine-tuning gives lower mean squared MMD than both independent training and angle-only fine-tuning across the displayed bandwidths. At $\sigma=1$, the reduction relative to independent training is approximately 25\%. This evaluation uses all coefficient orders, including those absent from the training loss.

Local refinement also brings the pretrained model's higher-order RMS magnitudes closer to the target in Fig.~\ref{fig:transfer}(b.ii), even though these orders do not enter the loss. In Fig.~\ref{fig:transfer}(c.ii), hypernetwork fine-tuning gives lower order-four coefficient error, while independent training gives lower order-eight error. Appendix~\ref{app:cdf-only} gives a separate CDF-KL-only comparison.

The lower distribution errors and reduced number of local updates make shared pretraining a promising starting point for preparing samplers at several Hamiltonian points. A trained family can supply these initializations before local data are used to refine each circuit.

\section{Larger systems}\label{sec:large-systems}
We now extend shared training and local refinement to a chain and a rectangular lattice with 1,024 measured qubits. Reference samples and classically estimated correlations allow us to train and compare these models without storing their full probability distributions.

\subsection{Training a 1,024-site chain}\label{sec:large-results}
The chain model uses a restricted Haldane circuit with 342 unmeasured ancillas and $r=2$ (Appendix~\ref{app:large-circuit}). At each Hamiltonian point, the hypernetwork supplies the 2,739 circuit angles using 87,693 weights.

The loss combines CDF-KL on 32 selected six-site marginals with the full-system Gaussian MMD of Eq.~\eqref{eq:mmd}, using sampled Fourier supports without an order cutoff. Exact gate cancellation further reduces the work needed for each coefficient while preserving the sampling circuit, as described in Appendix~\ref{app:mc-cancellation}. Appendix~\ref{app:large} gives the coefficient budgets and loss weights.

The fit uses nine initial reference points and 41 acquired points, with 20,000 reference samples per point and 1,500 updates in total. We evaluate the final checkpoint. The reference states are MPS approximations optimized with a bond-dimension cap of 128. MPS with limited bond dimension can retain useful phase features, as demonstrated for autonomous discovery in an XXZ chain~\cZhou, although the finite bond dimension can limit accuracy near transitions~\cTagliacozzo.

\subsection{Learned response peaks and local refinement}
We locate candidate phase boundaries using the mean susceptibility of the selected marginals,
\begin{equation}
\bar\chi_{22}=\frac1{4|\mathcal A|}\sum_{A\in\mathcal A}\sum_{\mathbf{z}}
\frac{[\partial_2 q_{A,\epsilon}(\mathbf{z})]^2}{q_{A,\epsilon}(\mathbf{z})},
\label{eq:localchi}
\end{equation}
where $q_{A,\epsilon}$ denotes the projected, smoothed marginal reconstructed from MC coefficient estimates. The MC random inputs are held fixed across the evaluation grid before differentiation. This indicator measures changes in the selected model marginals as $h_2/J$ varies.

Figure~\ref{fig:large}(a) recovers portions of both bulk phase-boundary branches from 50 reference points. The learned peaks follow segments of the bulk guides, with gaps and additional peaks elsewhere. This agreement shows that selected marginals can retain phase-boundary signatures at this system size.

At the marked point $(h_1/J,h_2/J)=(1.46,-1.40)$, Fig.~\ref{fig:large}(b) compares the pretrained circuit, its locally fine-tuned copy and an independently trained circuit against the same MPS reference. For both geometries, fine-tuning and independent training use the same 20,000 local samples and 600 updates. Evaluation uses separate held-out samples, with the black curve comparing the training and held-out sets from the same reference MPS.

The pretrained chain model already gives lower mean squared MMD than independent training at each of the ten tested bandwidths from $5$ to $64$, and fine-tuning reduces it further. At $\sigma=32$, the model errors remain above the train--test comparison. Fine-tuning gives $1.49\times10^{-5}$, compared with $4.48\times10^{-5}$ for independent training, an improvement by a factor of approximately three.

\subsection{A 1,024-site rectangular lattice}\label{sec:large-2d}
We also train on a 4 by 256 spin lattice with open boundaries, using the same circuit and training schedule as the chain. The CDF-KL marginals are now six-site rectangles. The rectangular Hamiltonian and reference calculations are specified in Appendix~\ref{app:large}. Figure~\ref{fig:large-2d}(a) shows several branches of peaks in the marginal susceptibility, identifying regions where the learned distributions change rapidly.

At the marked coordinate $(h_1/J,h_2/J)=(1.22,-0.04)$, local refinement gives a larger improvement than for the chain. The fine-tuned model has the lowest mean squared MMD of the three models at all 11 displayed bandwidths. At $\sigma=32$, it reaches $5.36\times10^{-6}$, compared with $3.98\times10^{-5}$ for independent training, a reduction by a factor of approximately seven. At this bandwidth, the fine-tuned error is also comparable in magnitude to the train--test comparison.

The pretrained lattice model has higher error than independent training at $\sigma=32$, yet becomes the more accurate model after fine-tuning. Shared pretraining therefore provides a useful starting point even when its direct prediction is less accurate. The improvements in both geometries extend the local-refinement results of Fig.~\ref{fig:transfer} to 1,024 measured qubits, using reference samples and classically estimated correlations.

\section{Discussion}\label{sec:discussion}
\paragraph{Shared classical training.}
We have shown that parameterised quantum circuits can learn families of ground-state measurement distributions through classical training. Each update combines losses from several Hamiltonian points, allowing the hypernetwork to learn how the circuit angles vary across the phase diagram. It can then predict angles at a new coordinate. When reference samples are available there, we can fine-tune the circuit for that particular distribution. Across 40 boundary points of the twelve-site chain, fine-tuning gave a lower final mean TVD than independent training.

The nine-site TFIM illustrates how the choice of circuit affects this fit. The hypernetwork IQP model and the RBM gave similar mean infidelities, $0.0115$ and $0.0106$, respectively. Adding three unmeasured ancillas to the IQP circuit reduced its mean infidelity to $0.00263$, about a fourfold improvement. Encoding the field through input rotations gave higher infidelity, even when a hypernetwork supplied the remaining angles. For this family, learning how the angles vary with the field worked better than the fixed input encoding we tested.

\paragraph{Scaling to larger systems.}
The Fourier representation let us train on the 1,024-site chain and the 4 by 256 lattice using selected correlations. We trained across 50 reference points per geometry and compared local fits at one point in each. At $\sigma=32$, fine-tuning gave about one third of the independent model's squared MMD on the chain and one seventh on the lattice. There, the shared model started with a higher error than the independently trained circuit and reached a lower error after local refinement.

\paragraph{Choice of objective.}
On the nine-site Haldane chain, MMD training gave lower mean infidelity across the parameter grid than CDF-KL when both used correlations through order four. The susceptibility of the CDF-KL fit traced portions of the phase boundary using the selected marginals. Infidelity measures agreement with the target at each point. Susceptibility measures how rapidly the learned distribution changes between nearby points. The position of a susceptibility peak can therefore agree with the target even when the probabilities on either side differ. The Fourier coefficients then show which correlations it reproduces and how their errors vary with order.

\paragraph{Information sharing and fine-tuning.}
Figure~\ref{fig:transfer}(b.i) shows that shared training already brings the low-order correlation magnitudes closer to the target than data-informed initialization, before any local fitting. Initializations that avoid barren plateaus can still lead to different minima~\cite{kulshrestha_exponentially_2026}. Shared training supplies a starting point informed by related distributions, and the hypernetwork parameterization can influence the minimum reached during local optimization~\cite{huang_spectral_2026}. In Fig.~\ref{fig:transfer}, the adaptively trained hypernetwork reached the independent fits' final mean TVD after about 550 local updates, compared with 1,500 for independent training. These counts exclude pretraining. With the same local update budget, hypernetwork fine-tuning gave lower mean squared MMD across the displayed bandwidths, including a reduction of about $25\%$ at $\sigma=1$.

This MMD comparison includes all coefficient orders, although the local loss uses only orders through four. Across the four selected initializations, fine-tuning brought the RMS coefficient magnitudes closer to the target at orders six, eight and ten. At order eight, however, independent training gave lower mean squared error for the individual coefficients. Agreement in the overall strength of the correlations can thus hide errors in the magnitude or sign of individual coefficients.

\paragraph{Reference acquisition and accuracy.}
Adaptive acquisition uses the correlations predicted by the current model to choose where to calculate the next ground state. In the nine-site fits, some of the acquired points cluster along branches of the learned boundary. At the same budget of 49 references, the fixed grid gave lower mean infidelity over the full diagram for both losses. When ground-state calculations are expensive, a practical use of the learned response is to choose a boundary point where a local fit is worth the cost. On the 1,024-site chain, the reference MPS at the evaluation point reached the sweep limit. The held-out samples quantify agreement with this finite-bond approximation to the ground state.

\paragraph{Dequantization.}
We draw reference samples directly from the MPS~\cFerris. For broader circuit families, estimating selected correlations can be easier than constructing a classical sampler for the full distribution~\cite{huang_spectral_2026}. The gate generators determine $r$ and hence the cost of estimating correlations during training. Even at a fixed $r$, which qubits the gates couple matters when seeking circuits whose outputs remain difficult to sample classically. Shared pretraining offers a promising route to avoiding dequantization if information learned across Hamiltonians helps the circuit recover correlations outside the local loss that are difficult for an efficient classical sampler to reproduce. Targets near phase boundaries, where correlations can change rapidly, provide a way to test this possibility by training both samplers on the same references and comparing their accuracy and sampling cost.

\section*{Code availability}
The code is available on \href{https://github.com/quantumsoftwarelab/HyperQCBM}{GitHub}.

\section*{Acknowledgements}
MHG proposed applying the train-on-classical, deploy-on-quantum approach~\cite{recio-armengol_train_2026a} to conditional QCBMs~\cite{kiss_conditional_2022} during the Spring School on Near-Term Quantum Computing at the Benasque Science Center. The hackathon challenge was proposed by A.~Cervera-Lierta, A.~P\'erez-Salinas, A.~Pozas-Kerstjens and J.~Arguello-Luengo. There, Guillermo Rubi\~nos-Rodr\'iguez demonstrated the usefulness of this approach through deployment on quantum hardware with error mitigation. MHG also thanks fellow team members Oriol Ball\'o and Daniel Quinn for discussions during the hackathon.

MHG is deeply grateful to Yuxuan Zhang for taking the time to explain automated phase-transition detection just before his train departed.

EK and MD acknowledge support from the Quantum Advantage TurboCHarger (QATCH) Programme. MHG was supported by the UK Engineering and Physical Sciences Research Council (EPSRC) through grant EP/W524384/1 and by the University of Edinburgh.

\section*{AI Usage}
The ideas in this work are original and were developed mainly through scientific discussions. We used GPT-6 Astra to assist with manuscript planning and writing, figure rendering and coding.

\clearpage
\onecolumngrid
\appendix
\makeatletter
\@addtoreset{figure}{section}
\@addtoreset{table}{section}
\makeatother
\renewcommand{\thefigure}{\Alph{section}\arabic{figure}}
\renewcommand{\thetable}{\Alph{section}\arabic{table}}
\renewcommand{\theHtable}{appendix.\Alph{section}.\arabic{table}}
\section{Targets, reference states and response calculation}\label{app:targets}
\subsection{Measurement basis and reference-state selection}
All probability comparisons use computational-basis outcomes of the measured qubits. The TFIM benchmark counts each undirected bond of the periodic $3\times3$ lattice once. The Haldane chain in Eq.~\eqref{eq:haldane} has open ends, with $n-2$ three-site terms and $n-1$ two-site terms. The normalization is in Pauli matrices, not spin operators $S^\alpha=\sigma^\alpha/2$.

At small-system degeneracies, we project a fixed seeded vector into the ground eigenspace and normalize it, using a second fixed vector if needed. The first vector is invariant under global bit flip. This fixes the reference state without separately fixing both sublattice parities. The choice matters because different states in a degenerate ground space can have different measurement probabilities. We omit undefined ground-state responses at degeneracies.

For a nondegenerate ground state, the component of its derivative orthogonal to the state is
\begin{equation}
\bigl(I-|\psi_0\rangle\langle\psi_0|\bigr)|\partial_\mu\psi_0\rangle=
\sum_{a>0}|\psi_a\rangle
\frac{\langle\psi_a|\partial_\mu H|\psi_0\rangle}{E_0-E_a}.
\end{equation}
It gives
$\partial_\mu p(\mathbf{x})=2\operatorname{Re}[\langle \mathbf{x}|\psi_0\rangle^*\langle \mathbf{x}|\partial_\mu\psi_0\rangle]$.
For learned distributions we differentiate through the circuit and the hypernetwork, holding trained parameters fixed. Input-rotation variants also include the explicit derivative of their encoding. Both calculations use the same dimensionless Hamiltonian coordinates in a comparison.

\subsection{From infidelity to susceptibility}\label{app:susceptibility}
Assume that $q_\la(\mathbf{x})$ is differentiable and its support remains fixed near $\la$, with positive probabilities on that support. All sums below are over this support. Normalization gives $\sum_{\mathbf{x}}\partial_\mu q(\mathbf{x})=0$. Expanding the overlap of the square-root probability vectors yields
\begin{equation}
\sum_{\mathbf{x}}\sqrt{q_\la(\mathbf{x})q_{\la+d\la}(\mathbf{x})}=1-\frac18\sum_{\mu\nu,\mathbf{x}}\frac{\partial_\mu q(\mathbf{x})\partial_\nu q(\mathbf{x})}{q(\mathbf{x})}d\lambda_\mu d\lambda_\nu+o(\|d\la\|^2).
\end{equation}
Squaring this overlap gives Eq.~\eqref{eq:chi}, connecting local distribution infidelity to susceptibility. This relation concerns nearby points within one distribution family. Small infidelity between a model and its target at individual points does not by itself guarantee accurate derivatives. At zero probabilities, the susceptibility expression requires a well-defined limit.

TVD does bound the discrepancy of a fixed bounded observable. If $|f(\mathbf{x})|\leq1$, then $|\E_pf-\E_qf|\leq2\operatorname{TVD}(p,q)$. Derivative and peak-position errors also depend on how the distributions vary between parameter points and on the shape of the peaks.

\subsection{Peak extraction and bulk guides}\label{app:peaks}
For the Haldane maps we inspect vertical slices at fixed $h_1/J$. Interior local maxima of $\chi_{22}$ are retained when their prominence is at least one tenth of the finite range on the slice. Undefined points break the slice into finite segments. Accepted maxima on adjacent columns are linked by mutual nearest neighbors only when their vertical separation is at most three grid rows. This preserves gaps and multiple branches without fitting the peaks to a reference curve.

The bulk guides in Fig.~\ref{fig:boundary} are the numerical boundaries published in Fig.~2(a) of Ref.~\cCong. They come from an infinite-system ground-state calculation. The exact reference peaks instead describe the finite-chain measurement distribution.

\section{Circuit families and model parameters}\label{app:gates}
The generator family determines the allowed circuit transformations and conserved Pauli products, summarized in Table~\ref{tab:gates}. We specify these families first, then give the finite gate lists and parameter counts used in the experiments.
\subsection{Compact Lie-algebra description}
Let $\mathcal P_{n_c}=\{I,X,Y,Z\}^{\otimes n_c}$ and let $N_A^J(P)$ count factors from $A\subset\{X,Y,Z\}$ on sites $J$. Write
$S_{\rm e}=\prod_{j\text{ even}}X_j$, $S_{\rm o}=\prod_{j\text{ odd}}X_j$, and $S_\alpha=\prod_j\sigma_j^\alpha$.
The Haldane family is
\begin{equation}
\mathfrak g_{\rm H}=\operatorname{span}_{\mathbb R}
\left\{iP:\begin{array}{l}
P\notin\{I,S_{\rm e},S_{\rm o},S_X\},\\
N_{\{Y,Z\}}^{\rm e}(P),N_{\{Y,Z\}}^{\rm o}(P)\text{ even}
\end{array}\right\}.
\label{eq:algebraH}
\end{equation}
We use the Pauli family proposed for the locally controlled Haldane chain in Ref.~\cBorn. For even $n_c\geq4$, the independently controlled Heisenberg family is
\begin{equation}
\mathfrak g_{\rm Heis}=\operatorname{span}_{\mathbb R}
\left\{iP:\begin{array}{l}
P\notin\{I,S_X,S_Y,S_Z\},\\
N_{\{X,Y\}}(P),N_{\{Y,Z\}}(P)\text{ even}
\end{array}\right\}.
\label{eq:algebraHeis}
\end{equation}
This Pauli characterization is used in~\cite{kökcü_fixed_2022,herrero-gonzalez_born_2025a}. Each commuting symmetry pair has four equally sized sectors. Removing its central directions gives
\begin{equation}
\mathfrak g_{\rm H}\cong\mathfrak g_{\rm Heis}
\cong\mathfrak{su}(2^{n_c-2})^{\oplus4},\qquad
\dim\mathfrak g=4^{n_c-1}-4,
\end{equation}
with different embeddings for the two families. The Haldane expression also applies for odd $n_c$ with both circuit sublattices nonempty. The Heisenberg expression above assumes even $n_c$.

For fermionic circuits, define Jordan--Wigner Majoranas
\begin{equation}
\gamma_{2j-1}=\left(\prod_{k<j}Z_k\right)X_j,\qquad
\gamma_{2j}=\left(\prod_{k<j}Z_k\right)Y_j.
\end{equation}
The real Lie span of $\gamma_a\gamma_b$, $a<b$, is $\mathfrak{so}(2n_c)$ with dimension $n_c(2n_c-1)$. Up to sign, its Hermitian rotation generators are $Z_j$ and
$A_i(\prod_{i<k<j}Z_k)B_j$, with $A,B\in\{X,Y\}$. The intervening $Z$ factors are part of the Jordan--Wigner representation. These quadratic generators produce fermionic Gaussian unitaries, also called fermionic linear optics, including when generators are repeated. With Gaussian inputs and computational-basis (occupation-number) measurements, their output distributions can be sampled efficiently on a classical computer~\cite{terhal_classical_2002,jozsa_matchgates_2008a}.

For the fermionic benchmarks, the computational zero state is Gaussian. The GHZ input is a superposition of two Gaussian occupation states whose relative phase can be tracked through the fermionic circuit~\cDias.

\begin{table*}[t]
\centering\begin{tabular}{lll}\toprule
Family & Permitted generators or selection rule & Conserved products\\\midrule
Haldane & Eq.~\eqref{eq:algebraH}, with even $Y/Z$ counts on $E$ and $O$ & $S_{\rm e},S_{\rm o}$\\
Heisenberg & Eq.~\eqref{eq:algebraHeis}, with even $X/Y$ and $Y/Z$ counts & $S_X,S_Z$\\
Fermionic & $Z_j$, $A_iZ_{i+1}\cdots Z_{j-1}B_j$, $A,B\in\{X,Y\}$ & $S_Z$\\
IQP & $X_S$ on selected nonempty subsets $S$ & all $X$ products\\
Restricted Haldane & $X_i$, allowed $X_iX_j$, $A_iB_j$ within $A_E$ or $A_O$, $A,B\in\{Y,Z\}$ & also $X_{\mathbf{x}}$ for $\mathbf{x}\in V^\perp$\\\bottomrule
\end{tabular}
\caption{\textbf{Gate families and conserved Pauli products.} $E$ and $O$ denote even and odd circuit sites, including ancillas. Each circuit uses a fixed ordered selection of the permitted generators. Gate weight counts nonidentity factors, independently of the distance between sites.}
\label{tab:gates}
\end{table*}

IQP generators commute and their Lie span is Abelian. With all nonempty $X$ strings through weight $w$, its dimension is $\sum_{k=1}^w\binom{n_c}{k}$. The Lie-algebra dimension describes the available generators. The circuit parameter count depends on the finite gate list chosen for each experiment.

\subsection{Finite gate lists and restrictions on active sites}
For Fig.~\ref{fig:haldane}, ancillas follow each group of three measured qubits. Haldane and Heisenberg words through weight three are sorted by increasing weight, support span and saved binary ordering. The first 298 distinct permitted words are retained. The fermionic basis has 276 words on twelve wires. Repeated rotations separated by noncommuting gates supply the remaining circuit angle parameters. IQP uses every one-, two- and three-site $X$ string, also giving 298 angles. Each model retains its fixed saved order. The GHZ preparation is on the complete register.

For the restricted Haldane family, choose nonempty qubit sets $A_E\subseteq E$ and $A_O\subseteq O$ within the two circuit sublattices (even and odd positions, respectively). The selected qubits are called active sites. The $Y/Z$ pair gates are permitted only within these sets, while the specified $X$ rotations are retained throughout the circuit. If the pairs connect each selected set, their binary incidence vectors give
\begin{equation}
r=(|A_E|-1)+(|A_O|-1).
\label{eq:rankactive}
\end{equation}
More generally, a graph of allowed pairs with $c$ connected components, including isolated sites, gives $r=n_c-c$ because each connected component spans the even-parity labels on its vertices. Adding a new pair inside a connected component does not increase $r$. All-$X$ words have zero vector $\mathbf{z}$ and also leave $r$ unchanged.

The unrestricted twelve-wire Haldane list has $r=10$. The $r=4$ and $r=2$ comparisons retain 298 parametrized rotations by allowed repetitions, but their additional conserved $X$ products change the model. Each adaptive fit also chooses its own reference coordinates, so the comparison changes both the allowed gates and the training data. Figure~\ref{fig:transfer} instead uses fifteen wires, three active sites on each parity and 144 distinct permitted generators, repeated to obtain 316 parametrized rotations. Appendix~\ref{app:large-circuit} specifies the circuit for both 1,024-site geometries.

\subsection{Hypernetwork and RBM counts}
For input dimension $d_\lambda$, a single tanh hidden layer of width $w$ and an output of length $2d$ give
\begin{align}
h(\la)&=\tanh(W_1\la+b_1),\\
(\gamma_\phiv,\beta_\phiv)&=W_2h+b_2,\\
N_{\rm hyper}&=w(d_\lambda+1)+2d(w+1).
\end{align}
The $d$ base angles are additional trainable parameters. Fixed scaling of the network output does not add weights. This yields 9,581 hypernetwork weights for Fig.~\ref{fig:haldane}, 10,157 for Fig.~\ref{fig:transfer}, and 87,693 for Fig.~\ref{fig:large}, each with $w=15$ and $d_\lambda=2$.

The RBM retains a shared visible-hidden interaction matrix and changes the visible and hidden biases using the HyperRBM rule~\cTonner. With visible bits $\mathbf{x}$ and binary hidden units, write
\begin{equation}
f_\la(\mathbf{x})=\exp(a_\la\cdot \mathbf{x})\prod_j[1+\exp(b_{\la,j}+W_j\cdot \mathbf{x})].
\end{equation}
The symmetrized unnormalized weight is $f_\la(\mathbf{x})+f_\la(\bar{\mathbf{x}})$. We normalize it by summing the weights of every possible outcome for the reported small systems. The Haldane comparison uses 29 hidden units. Its 261 interaction weights and 38 biases give 299 generator parameters. A width-122 network for the 38 biases has 9,714 weights. The interaction matrix is shared across Hamiltonian points and trained jointly with the bias network.

\section{Derivation and special cases of the Monte Carlo estimator}\label{app:mc}
We derive the coefficient estimator introduced in Sec.~\ref{sec:mc}, establish its variance and computational cost, and give the IQP and GHZ special cases. The derivation retains the order of the Pauli gates and applies to fixed circuit parameters. Appendix~\ref{app:losses} explains how coefficient estimates enter the training losses.
\begin{proposition}[Classical estimation for ordered Pauli circuits]\label{prop:mc-estimator}
Let $U_\tht=e^{-i\theta_dP_d}\cdots e^{-i\theta_1P_1}$ be a circuit of $d\geq1$ Pauli rotations with Hermitian generators on $n_c$ qubits, with fixed real angles and input $|\mathbf0\rangle=|0\rangle^{\otimes n_c}$. Write $P_j=i^{\nu_j}X^{\mathbf{x}_j}Z^{\mathbf{z}_j}$ as in Sec.~\ref{sec:mc}, and let
\[
r=\dim_{\mathbb F_2}\operatorname{span}\{\mathbf{z}_1,\ldots,\mathbf{z}_d\}.
\]
For any subset $S$ of the measured qubits, set $c_S=\langle\mathbf0|U_\tht^\dagger Z_SU_\tht|\mathbf0\rangle$. The estimator in Eq.~\eqref{eq:estimator}, using $N_{\rm MC}$ independent uniform bitstrings, satisfies
\begin{equation}
\E\widehat c_S=c_S,\qquad
\Var(\widehat c_S)\leq\frac{1-c_S^2}{N_{\rm MC}},\qquad
\Pr\!\left(|\widehat c_S-c_S|\geq\varepsilon\right)
\leq2e^{-N_{\rm MC}\varepsilon^2/2}.
\label{eq:mc-proposition-bounds}
\end{equation}
Consequently, for $0<\varepsilon\leq1$ and $0<\delta<1$, taking
\[
N_{\rm MC}=\left\lceil2\varepsilon^{-2}\log(2/\delta)\right\rceil
\]
suffices for additive error $\varepsilon$ with probability at least $1-\delta$.

Let $W_x=\sum_{j=1}^d|\operatorname{supp}(\mathbf{x}_j)|\leq dn_c$. Including preprocessing, a direct implementation uses
\begin{align}
T&=O\!\left(dn_c^2+d(1+r)2^r+N_{\rm MC}[n_c+W_x+d2^r]\right),\label{eq:mc-total-time}\\
\mathcal M&=O\!\left(dn_c+d(1+r)2^r\right)\label{eq:mc-total-storage}
\end{align}
operations and storage, respectively. Storage includes the circuit description, precomputed gate tables and working vectors, and is independent of $N_{\rm MC}$ when samples are processed sequentially. These counts treat numerical arithmetic and array indexing as unit-cost operations, with gate sines and cosines supplied as numerical inputs. Unbiasedness and the statistical bounds assume exact gate updates and do not include numerical roundoff. If $d=\operatorname{poly}(n_c)$ and $r=O(\log n_c)$, the operation count is polynomial in $n_c$, $\varepsilon^{-1}$ and $\log(\delta^{-1})$, and the storage is polynomial in $n_c$.
\end{proposition}

The proof of Proposition~\ref{prop:mc-estimator} begins in Appendix~\ref{app:mc-update}, where we show that the mean overlap of two vectors with $2^r$ amplitudes equals the Fourier coefficient $c_S$. Appendix~\ref{app:mc-cost} completes the proof by bounding the variance, the number of draws needed for a given accuracy, and the time and storage required.
\subsection{Invariant sets of bitstrings}\label{app:mc-update}
\paragraph*{Proof of Proposition~\ref{prop:mc-estimator}.}
Use $|\widetilde{\mathbf{b}}\rangle=H^{\otimes n_c}|\mathbf{b}\rangle$ for the $X$ basis. On one site, $X_i|\widetilde{\mathbf{b}}\rangle=(-1)^{b_i}|\widetilde{\mathbf{b}}\rangle$, while $Z_i$ flips $b_i$. Thus
\begin{align}
Z^{\mathbf{z}_j}|\widetilde{\mathbf{b}}\rangle
 &=|\widetilde{\mathbf{b}\oplus\mathbf{z}_j}\rangle,\\
X^{\mathbf{x}_j}|\widetilde{\mathbf{b}\oplus\mathbf{z}_j}\rangle
 &=(-1)^{\mathbf{x}_j\cdot(\mathbf{b}\oplus\mathbf{z}_j)}
 |\widetilde{\mathbf{b}\oplus\mathbf{z}_j}\rangle.
\end{align}
Applying the rightmost factor first in $P_j=i^{\nu_j}X^{\mathbf{x}_j}Z^{\mathbf{z}_j}$ gives
\begin{equation}
P_j|\widetilde{\mathbf{b}}\rangle=i^{\nu_j}(-1)^{\mathbf{x}_j\cdot(\mathbf{b}\oplus \mathbf{z}_j)}
|\widetilde{\mathbf{b}\oplus \mathbf{z}_j}\rangle,
\end{equation}
Since $P_j^2=I$, its rotation is $e^{-i\theta_jP_j}=\cos\theta_j I-i\sin\theta_jP_j$. It mixes only the two bitstrings separated by $\mathbf{z}_j$. Repeating these shifts preserves the span of basis states with bitstrings in $\mathbf{b}+V$. Use the basis matrix $B$ from Sec.~\ref{sec:mc}. The binary vector $\boldsymbol{\ell}_j\in\mathbb F_2^r$ is defined by $B\boldsymbol{\ell}_j=\mathbf{z}_j$. It specifies which vector entries gate $j$ mixes. The binary vector $\mathbf{h}_j=B^{\mathsf T}\mathbf{x}_j\in\mathbb F_2^r$ specifies their relative signs. For zero input, $|\mathbf0\rangle=2^{-n_c/2}\sum_{\mathbf{b}}|\widetilde{\mathbf{b}}\rangle$. A coset contains $2^r$ terms and has squared norm $2^{r-n_c}$. Dividing its amplitudes by the square root of that weight gives the normalized initial vector
$w_0^{(\mathbf{b})}(\mathbf{t})=2^{-r/2}$, $\mathbf{t}\in\mathbb F_2^r$. Its exact gate update is
\begin{align}
w_j^{(\mathbf{b})}(\mathbf{t})={}&\cos\theta_j\,w_{j-1}^{(\mathbf{b})}(\mathbf{t})\nonumber\\
&-i^{\nu_j+1}\sin\theta_j(-1)^{\mathbf{x}_j\cdot \mathbf{b}+\mathbf{h}_j\cdot \mathbf{t}}
 w_{j-1}^{(\mathbf{b})}(\mathbf{t}\oplus\boldsymbol{\ell}_j).
\label{eq:vectorupdate}
\end{align}
To check the phase, the source bitstring for the second term is $\mathbf{b}\oplus B\mathbf{t}\oplus\mathbf{z}_j$. Substitution in the Pauli action above gives $i^{\nu_j}(-1)^{\mathbf{x}_j\cdot\mathbf{b}+\mathbf{h}_j\cdot\mathbf{t}}$, which is multiplied by $-i\sin\theta_j$. Both terms use the pre-gate vector. The update is unitary within the invariant set, so the vector remains normalized. Bitstring arithmetic is modulo two, while $\nu_j$ is an integer phase exponent. The representative for the second vector is $\mathbf{b}\oplus\mathbf{s}$, and the same coordinate $\mathbf{t}$ pairs bitstrings shifted by $\mathbf{s}$ in Eq.~\eqref{eq:estimator}. This alignment is required even when the two bitstrings belong to the same coset. Replacing the shifted representative by a canonical one without permuting its entries would change the estimator.

There are $K_V=2^{n_c-r}$ distinct sets. The zero input $|\mathbf0\rangle=|0\rangle^{\otimes n_c}$ has equal weight in all of them. Its final amplitude at $\mathbf{b}\oplus B\mathbf{t}$ is
\begin{equation}
\langle\widetilde{\mathbf{b}\oplus B\mathbf{t}}|U_\tht|\mathbf0\rangle=K_V^{-1/2}w_d^{(\mathbf{b})}(\mathbf{t}).
\end{equation}
The observable $Z_{\mathbf{s}}$ shifts $X$-basis bitstrings by $\mathbf{s}$. Hence, if $\mathcal R$ contains one representative of each set,
\begin{equation}
c_S=\frac1{K_V}\sum_{\mathbf{b}\in\mathcal R}
\operatorname{Re}\sum_{\mathbf{t}} w_d^{(\mathbf{b}\oplus \mathbf{s})}(\mathbf{t})^*w_d^{(\mathbf{b})}(\mathbf{t}).
\end{equation}
Uniform bitstrings sample these sets uniformly. More explicitly, replacing $\mathbf{b}$ by $\mathbf{b}\oplus B\mathbf{u}$ sends each vector entry $\mathbf{t}$ to $\mathbf{t}\oplus\mathbf{u}$ in both vectors. Their overlap is unchanged. Thus every coset contributes the same sample value for each of its $2^r$ representatives, allowing us to sample bitstrings without enumerating the cosets. This proves the expectation identity in Eq.~\eqref{eq:variance}.

\subsection{Variance and computational cost}\label{app:mc-cost}
By Cauchy--Schwarz, $|G_S|\leq1$. Independent draws therefore obey
\begin{equation}
\Var(\widehat c_S)=\frac{K_V^{-1}\sum_{\mathbf{b}\in\mathcal R}G_S(\mathbf{b})^2-c_S^2}{N_{\rm MC}}
\leq\frac{1-c_S^2}{N_{\rm MC}}\leq\frac1{N_{\rm MC}}.
\end{equation}
Hoeffding's bound gives
\begin{equation}
\Pr(|\widehat c_S-c_S|\geq\varepsilon)
\leq2\exp(-N_{\rm MC}\varepsilon^2/2).
\end{equation}
For a fixed collection of $M$ coefficients, a union bound replaces the prefactor by $2M$. Independence between different coefficients is not needed for that bound. These bounds hold for fixed circuit parameters and independent estimator draws.

To account for preprocessing, binary elimination finds a basis $B$ and the coordinates $\boldsymbol{\ell}_j$ in $O(dn_c^2)$ operations. Forming $\mathbf h_j=B^{\mathsf T}\mathbf x_j$ and the partner offsets $\mathbf x_j\cdot\mathbf s$ also fits within this bound. For each gate, tabulate the permutation $\mathbf t\mapsto\mathbf t\oplus\boldsymbol\ell_j$ and signs $(-1)^{\mathbf h_j\cdot\mathbf t}$ in Eq.~\eqref{eq:vectorupdate}. Direct construction takes $O(d(1+r)2^r)$ binary operations and bits of storage, including the $r$-bit permutation entries. The circuit description and binary basis data require $O(dn_c)$ further storage, since $r\leq\min(d,n_c)$.

Each draw uses two vectors of length $2^r$, each passed through $d$ gates. Generating its $n_c$ random bits costs $O(n_c)$, and calculating the parities $\mathbf{x}_j\cdot\mathbf{b}$ costs $O(W_x)$ by visiting the nonzero entries of each $\mathbf{x}_j$. The partner parities follow from $\mathbf{x}_j\cdot(\mathbf{b}\oplus\mathbf{s})=\mathbf{x}_j\cdot\mathbf{b}+\mathbf{x}_j\cdot\mathbf{s}$. With the tables available, each gate update takes $O(2^r)$ arithmetic and indexing operations. The final overlap has the same order of cost. Processing the draws sequentially therefore takes $O(N_{\rm MC}[n_c+W_x+d2^r])$ operations and $O(2^r)$ numerical working storage, including temporary vectors. Only a running sum is needed. Adding preprocessing gives Eqs.~\eqref{eq:mc-total-time} and~\eqref{eq:mc-total-storage}. Substituting the stated sample count and $r=O(\log n_c)$ gives the polynomial bounds.\hfill$\square$

The empty circuit can be evaluated directly. Finite-precision evaluation introduces an additional numerical error, separate from the sampling error in Eq.~\eqref{eq:mc-proposition-bounds}. Reverse-mode differentiation can require more memory unless intermediate states are recomputed.
The estimator is a finite sum of differentiable functions of the angles, and its sampling distribution is angle-independent. Thus expectation and differentiation can be interchanged for its coefficient gradient. This does not give the same variance bound for that gradient, or remove bias from a subsequent nonlinear loss. At an adaptive training step, the fixed-angle statement applies conditionally when fresh bitstrings are sampled independently of the current parameters. It does not justify evaluating a fitted model on the same random bitstrings used to select it.

Enumerating all supported cosets removes MC variance. For the full Haldane gate list, however, each amplitude vector has exponentially many entries, so even a small number of cosets can be costly. Restricting the active sites reduces this cost, as used in Figs.~\ref{fig:restrictions}--\ref{fig:large-2d}. Other simulation methods may remain efficient for gate sets requiring larger vectors in this estimator.

\subsection{IQP phase estimator}\label{app:mc-iqp}
For $P_j=X^{\mathbf{x}_j}$ every vector $\mathbf{z}_j$ vanishes, so $r=0$ and each vector has one entry, initially equal to one. The gate acts diagonally,
\begin{equation}
w_j^{(\mathbf{b})}=e^{-i\theta_j(-1)^{\mathbf{x}_j\cdot\mathbf{b}}}w_{j-1}^{(\mathbf{b})}.
\end{equation}
Multiplying these factors gives $w_\tht^{(\mathbf{b})}=e^{-i\Phi_\tht(\mathbf{b})}$, where $\Phi_\tht(\mathbf{b})=\sum_j\theta_j(-1)^{\mathbf{x}_j\cdot\mathbf{b}}$. Substitution in the overlap yields
\begin{align}
G_S(\mathbf{b};\tht)
 &=\operatorname{Re}\left[e^{i\Phi_\tht(\mathbf{b}\oplus\mathbf{s})}e^{-i\Phi_\tht(\mathbf{b})}\right]\\
 &=\cos[\Phi_\tht(\mathbf{b}\oplus\mathbf{s})-\Phi_\tht(\mathbf{b})].
\end{align}
Reducing the phase difference term by term gives
\begin{align}
\Phi_\tht(\mathbf{b}\oplus\mathbf{s})-\Phi_\tht(\mathbf{b})
 &=\sum_j\theta_j(-1)^{\mathbf{x}_j\cdot\mathbf{b}}
 \left[(-1)^{\mathbf{x}_j\cdot\mathbf{s}}-1\right]\\
 &=-2\sum_{j:\,\mathbf{x}_j\cdot\mathbf{s}=1}
 \theta_j(-1)^{\mathbf{x}_j\cdot\mathbf{b}}.
\end{align}
Terms with $\mathbf{x}_j\cdot\mathbf{s}=0$ cancel. Since cosine is even, the remaining expression gives the sample used in Eq.~\eqref{eq:iqp-recovery}. Averaging it gives
\begin{equation}
c_S=\E_{\mathbf{b}}\cos[\Phi_\tht(\mathbf{b}\oplus \mathbf{s})-\Phi_\tht(\mathbf{b})].
\label{eq:iqpsample}
\end{equation}
This is the phase-average form underlying classical IQP optimization~\cite{nest_simulating_2010,armengol_iqpopt_2026}. With zero input, $\mathbf{b}$ is uniform over all bitstrings. Appendix~\ref{app:mc-ghz} derives the sampling distribution for GHZ input.

\subsection{GHZ input and even-parity sampling}\label{app:mc-ghz}
\begin{corollary}[GHZ input with parity-preserving gates]\label{cor:mc-ghz}
Replace the zero input in Proposition~\ref{prop:mc-estimator} by $|\mathrm{GHZ}+\rangle=(|0\rangle^{\otimes n_c}+|1\rangle^{\otimes n_c})/\sqrt2$, and suppose $\mathbf1\cdot\mathbf z_j=0$ for every gate. For even $|S|$, the same estimator, with independent bitstrings drawn uniformly from the even-parity subspace, obeys the same expectation, variance, probability and complexity bounds. For odd $|S|$, $c_S=0$ exactly.
\end{corollary}

\paragraph*{Proof.}
The two computational-basis components have $X$-basis expansions
\begin{align}
|0\rangle^{\otimes n_c}&=2^{-n_c/2}\sum_{\mathbf{b}}|\widetilde{\mathbf{b}}\rangle,\\
|1\rangle^{\otimes n_c}&=2^{-n_c/2}\sum_{\mathbf{b}}(-1)^{\mathbf1\cdot\mathbf{b}}|\widetilde{\mathbf{b}}\rangle.
\end{align}
Adding them and dividing by $\sqrt2$ gives amplitude $2^{-(n_c+1)/2}[1+(-1)^{\mathbf1\cdot\mathbf{b}}]$. This vanishes for odd-parity bitstrings and equals $2^{-(n_c-1)/2}$ for even-parity bitstrings. Therefore
\begin{equation}
|\mathrm{GHZ}+\rangle=2^{-(n_c-1)/2}
\sum_{\mathbf{b}:\,\mathbf1\cdot \mathbf{b}=0}|\widetilde{\mathbf{b}}\rangle.
\end{equation}
When every $\mathbf{z}_j$ has even parity, $V$ lies in the even-parity subspace, and each coset starting at an even-parity bitstring stays inside this support. Dividing the $2^{n_c-1}$ supported bitstrings into sets of size $2^r$ gives $K_{\rm GHZ}=2^{n_c-r-1}$ supported cosets. Each has weight $K_{\rm GHZ}^{-1}$ and the same normalized initial entries $2^{-r/2}$ as before. Thus the gate update is unchanged. For even $|S|$, $\mathbf1\cdot\mathbf{s}=0$, so the partner coset at $\mathbf{b}\oplus\mathbf{s}$ is also supported, and
\begin{equation}
c_S=\frac1{K_{\rm GHZ}}\sum_{\mathbf{b}\in\mathcal R_{\rm even}}G_S(\mathbf{b};\tht),
\end{equation}
where $\mathcal R_{\rm even}$ contains one representative of each supported coset. Sampling uniformly among even-parity bitstrings implements this average because every coset has the same size. The normalization and variance proof in Appendix~\ref{app:mc-cost} therefore still apply. For odd $|S|$, $Z_{\mathbf{s}}$ shifts into the unoccupied odd-parity sector, whose overlap with the state is zero. Hence $c_S=0$. Sampling $n_c-1$ independent fair bits and choosing the final bit to make the total parity even gives a uniform supported bitstring in $O(n_c)$ operations. The gate tables and vector updates are unchanged, so the complexity bounds also apply.\hfill$\square$

For the Haldane and IQP circuits with GHZ input used here, only even-order coefficients therefore need MC evaluation. Other inputs or parity-changing gates require their corresponding amplitudes and sampling weights.

\subsection{Spans of full and restricted Haldane gate vectors}\label{app:mc-haldane}
Each $Y/Z$ pair on sites $u,v$ has binary vector $\mathbf{e}_u\oplus\mathbf{e}_v$, independently of whether its factors are $Y$ or $Z$. Here $\mathbf{e}_u$ has a one at site $u$ and zeros elsewhere. An $X$-type gate has zero binary vector. Form a graph whose vertices are circuit sites and whose edges are the permitted $Y/Z$ pairs. An edge vector has even parity within its connected component, and so does every binary sum of edge vectors.

Conversely, consider a connected component $C$ and fix a root site $u\in C$. For every other site $v\in C$, a path $u=v_0,v_1,\ldots,v_m=v$ gives
\begin{equation}
\bigoplus_{k=0}^{m-1}(\mathbf{e}_{v_k}\oplus\mathbf{e}_{v_{k+1}})
=\mathbf{e}_u\oplus\mathbf{e}_v,
\end{equation}
because each internal vertex occurs twice. The $|C|-1$ root-pair vectors are linearly independent because the coordinate at $v\ne u$ occurs in just its own vector. They also span every even-parity bitstring $\mathbf{z}$ supported on $C$, since
\begin{equation}
\mathbf{z}=\bigoplus_{\substack{v\in C\setminus\{u\}\\z_v=1}}
(\mathbf{e}_u\oplus\mathbf{e}_v).
\end{equation}
The root coordinate agrees because $\mathbf{z}$ has even parity. Thus the span of these vectors on $C$ has dimension $|C|-1$. A singleton contributes zero. Components have disjoint supports, so a graph with $c$ components has total dimension $\sum_C(|C|-1)=n_c-c$.

For the full Haldane pair set, the even and odd circuit sublattices are each connected and nonempty. There are two components, giving
\begin{equation}
V_{\rm H}=\{\mathbf{z}:\mathbf{z}\cdot\mathbf1_E=\mathbf{z}\cdot\mathbf1_O=0\},
\qquad r=(|E|-1)+(|O|-1)=n_c-2.
\end{equation}
Additional Haldane gate vectors with even parity on each sublattice already belong to this span and do not enlarge it. Each simulated amplitude vector has $2^{n_c-2}$ entries.

For the restricted circuit, the permitted pairs connect only the nonempty sets $A_E$ and $A_O$. Every site outside those sets is an isolated vertex, so
\begin{equation}
c=2+n_c-|A_E|-|A_O|,\qquad
r=n_c-c=|A_E|+|A_O|-2.
\end{equation}
This proves Eq.~\eqref{eq:rank-examples}. Two sites in each selected set give $r=2$ and four entries. Three give $r=4$ and sixteen entries. Choosing a total of $O(\log n_c)$ selected sites makes $2^r$ polynomial in $n_c$. The complexity statement also requires the gate count and requested accuracy to scale as specified in Appendix~\ref{app:mc-cost}. If a selected set is disconnected, count its separate components instead of applying the connected-set formula.

\subsection{Two-entry updates and a noncommuting Haldane example}\label{app:mc-noncommuting}
Let $i,j$ be distinct sites on the same circuit sublattice, and retain $Z_iZ_j$ together with $X$-type generators. The only nonzero gate vector is $\mathbf{z}=\mathbf{e}_i\oplus\mathbf{e}_j$, so $V=\{\mathbf0,\mathbf{z}\}$ and $r=1$. Choose $B=\mathbf{z}$. The two entries, indexed by $t=0,1$, correspond to bitstrings $\mathbf{b}$ and $\mathbf{b}\oplus\mathbf{z}$. On these basis states, $Z_iZ_j$ interchanges the entries. Expanding its exponential therefore gives
\begin{equation}
R(\beta)=e^{-i\beta\left(\begin{smallmatrix}0&1\\1&0\end{smallmatrix}\right)}
=\begin{pmatrix}\cos\beta&-i\sin\beta\\-i\sin\beta&\cos\beta\end{pmatrix}.
\end{equation}
An $X^{\mathbf{x}}$ rotation is diagonal, with entry $t$ multiplied by $e^{-i\alpha(-1)^{\mathbf{x}\cdot(\mathbf{b}\oplus t\mathbf{z})}}$. In particular, put $\xi=(-1)^{b_i}$. The two eigenvalues of $X_i$ are $\xi$ and $-\xi$, giving
\begin{equation}
D(\alpha;\xi)=\begin{pmatrix}e^{-i\alpha\xi}&0\\0&e^{i\alpha\xi}\end{pmatrix}.
\end{equation}
These are the two-entry updates stated in the main text. Starting from $w_0=(1,1)^{\mathsf T}/\sqrt2$, apply them in circuit order. For $U=e^{-i\beta Z_iZ_j}e^{-i\alpha X_i}$,
\begin{equation}
w_1=D(\alpha;\xi)w_0=\frac1{\sqrt2}
\begin{pmatrix}e^{-i\alpha\xi}\\e^{i\alpha\xi}\end{pmatrix},
\qquad w_2=R(\beta)w_1.
\end{equation}
Multiplication gives
\begin{equation}
w^{(\mathbf{b})}=\frac1{\sqrt2}
\begin{pmatrix}
\cos\beta\,e^{-i\alpha\xi}-i\sin\beta\,e^{i\alpha\xi}\\
\cos\beta\,e^{i\alpha\xi}-i\sin\beta\,e^{-i\alpha\xi}
\end{pmatrix}.
\end{equation}
A correlator sample is the real overlap of this vector with the one at $\mathbf{b}\oplus\mathbf{s}$, pairing the same $t$ in both vectors as in Eq.~\eqref{eq:estimator}. The final $Z_iZ_j$ rotation commutes with every measured $Z_S$, so
\begin{equation}
U^\dagger Z_SU
=e^{i\alpha X_i}e^{i\beta Z_iZ_j}Z_Se^{-i\beta Z_iZ_j}e^{-i\alpha X_i}
=e^{i\alpha X_i}Z_Se^{-i\alpha X_i}.
\end{equation}
Thus all these correlators are independent of $\beta$, despite the noncommuting gates. To obtain dependence on the intervening rotation, take zero input and append an $X_i$ rotation,
\begin{equation}
U=e^{-i\gamma X_i}e^{-i\beta Z_iZ_j}e^{-i\alpha X_i}.
\end{equation}
Its gate vectors still span a one-dimensional binary space, so $r=1$. The final vector is $w_3=D(\gamma;\xi)R(\beta)D(\alpha;\xi)w_0$. Write it as $w_3=2^{-1/2}(f_0(\xi),f_1(\xi))^{\mathsf T}$, where
\begin{align}
f_0(\xi)&=e^{-i\gamma\xi}\left(\cos\beta\,e^{-i\alpha\xi}-i\sin\beta\,e^{i\alpha\xi}\right),\\
f_1(\xi)&=e^{i\gamma\xi}\left(\cos\beta\,e^{i\alpha\xi}-i\sin\beta\,e^{-i\alpha\xi}\right).
\end{align}
For $Z_i$, the partner bitstring is $\mathbf{b}\oplus\mathbf{e}_i$, which replaces $\xi$ by $-\xi$ and interchanges $f_0$ and $f_1$. The sample is consequently
\begin{equation}
G_{\{i\}}(\mathbf{b})=\frac12\operatorname{Re}
\left[f_1(\xi)^*f_0(\xi)+f_0(\xi)^*f_1(\xi)\right]
=\operatorname{Re}[f_0(\xi)^*f_1(\xi)].
\end{equation}
The product simplifies to
\begin{align}
f_0(\xi)^*f_1(\xi)
 &=e^{2i\gamma\xi}\left(\cos^2\beta\,e^{2i\alpha\xi}
                         +\sin^2\beta\,e^{-2i\alpha\xi}\right)\\
 &=e^{2i\gamma\xi}\left[\cos(2\alpha)+i\xi\cos(2\beta)\sin(2\alpha)\right].
\end{align}
Taking the real part and using $\xi^2=1$ gives a value independent of the sampled bitstring. Its average is therefore
\begin{equation}
\begin{split}
\langle Z_i\rangle={}&\cos(2\gamma)\cos(2\alpha)\\
&-\sin(2\gamma)\sin(2\alpha)\cos(2\beta).
\end{split}
\end{equation}
The same result follows by conjugating the observable, starting with the last applied gate. For anticommuting Pauli words, $e^{i\theta P}Qe^{-i\theta P}=Q\cos(2\theta)+iPQ\sin(2\theta)$. Here $iX_iZ_i=Y_i$, $iZ_iZ_jY_i=X_iZ_j$ and $iX_iY_i=-Z_i$. The successive observables are
\begin{align}
Q_1&=e^{i\gamma X_i}Z_ie^{-i\gamma X_i}
    =Z_i\cos(2\gamma)+Y_i\sin(2\gamma),\\
Q_2&=e^{i\beta Z_iZ_j}Q_1e^{-i\beta Z_iZ_j}\nonumber\\
   &=Z_i\cos(2\gamma)+Y_i\sin(2\gamma)\cos(2\beta)
     +X_iZ_j\sin(2\gamma)\sin(2\beta),\\
Q_3&=e^{i\alpha X_i}Q_2e^{-i\alpha X_i}\nonumber\\
   &=Z_i[\cos(2\gamma)\cos(2\alpha)-\sin(2\gamma)\cos(2\beta)\sin(2\alpha)]\nonumber\\
   &\quad+Y_i[\cos(2\gamma)\sin(2\alpha)+\sin(2\gamma)\cos(2\beta)\cos(2\alpha)]\nonumber\\
   &\quad+X_iZ_j\sin(2\gamma)\sin(2\beta).
\end{align}
On $|0\rangle^{\otimes n_c}$, the expectations of $Z_i,Y_i,X_iZ_j$ are $1,0,0$, respectively. The coefficient of $Z_i$ in $Q_3=U^\dagger Z_iU$ reproduces the overlap result. The example shows how noncommuting gates can affect the measured coefficient while requiring only two amplitudes per vector. For the parity-preserving GHZ input, this odd-order coefficient vanishes by Appendix~\ref{app:mc-ghz}.

\subsection{Exact gate cancellation for selected observables}\label{app:mc-cancellation}
Some gates cancel in $U^\dagger Z_{\mathbf{s}}U$ because they commute with the complete backward-evolved operator at their location. For a Pauli term $Q$,
\begin{equation}
e^{i\theta P}Qe^{-i\theta P}=
\begin{cases}
Q,&[P,Q]=0,\\
Q\cos2\theta+iPQ\sin2\theta,&\{P,Q\}=0.
\end{cases}
\end{equation}
The implementation keeps a binary span containing the Pauli labels of the backward-evolved observable. A gate commuting with every basis element of that span commutes with the complete observable and can be removed. When a gate anticommutes with any basis element, it is retained and its Pauli label is added to the span. This rule removes gates without changing the selected coefficient, although it may retain some gates whose effects cancel at particular angles. The estimator uses the remaining gates in their original order. The full circuit, including gates omitted from this coefficient calculation, is used for sampling.

\section{Losses from the available coefficients}\label{app:losses}
This appendix gives the marginal reconstruction used by CDF-KL and the independent coefficient estimates used by MMD. Both losses are evaluated at each reference point and averaged to train the shared model.
\subsection{Marginal reconstruction and valid cumulative probabilities}
Let $\widehat q_A$ be the vector obtained from Eq.~\eqref{eq:marginal} using estimated coefficients. Its entries sum to one if $c_\varnothing=1$, but may be negative. We compute its Euclidean simplex projection
\begin{equation}
\Pi_\Delta(v)=\arg\min_{u_i\geq0,\,\sum_i u_i=1}\|u-v\|_2^2.
\end{equation}
Explicitly, $u_i=\max(v_i-\tau,0)$, with $\tau$ chosen to make the sum one. Sorting the entries determines the active set and $\tau$. The probability used for cumulative sums is
\begin{equation}
q_{A,\epsilon}=(1-\epsilon)\Pi_\Delta(\widehat q_A)+\epsilon\,2^{-|A|}\mathbf1.
\end{equation}
The same operation is applied to the empirical target marginal. Projection leaves an exact normalized nonnegative vector unchanged before smoothing. For $0<\epsilon<1$, every outcome has probability at least $\epsilon 2^{-\ell}$, where $\ell=|A|$. At each retained threshold, the cumulative probability therefore lies in $[\epsilon 2^{-\ell},1-\epsilon 2^{-\ell}]$. The Bernoulli KL terms and their derivatives with respect to the cumulative probabilities are then finite. The projection itself remains piecewise differentiable. We use $\epsilon=10^{-3}$ in the reported restricted-loss experiments. The CDF omits the last threshold, whose probability is identically one.

With all thresholds and fixed ordering, zero CDF-KL implies equality of the smoothed marginal vectors because consecutive CDF differences recover their individual probabilities. Equality of selected marginals can still leave the full distributions different. Each $\ell$-site marginal requires all $2^\ell$ coefficients on that subset, even when only some subsets are used. The small-system four-site studies use the union of all such subsets. The large-system study selects only 32 windows.

The bit order in each marginal is fixed throughout training. Changing it changes the cumulative events and can change CDF-KL. The quadratic expansion
\begin{equation}
d_{\rm B}(u,u+\delta)=\frac{\delta^2}{2u(1-u)}+O(\delta^3)
\end{equation}
shows why the cumulative events are weighted differently from direct squared coefficient errors. Projection and logarithms generally bias the MC loss and its gradient relative to CDF-KL evaluated with exact coefficients and the same smoothing. The smoothing parameter separately controls how much each marginal is mixed with the uniform distribution.

Each projected marginal is a valid probability vector, but independently projected marginals can disagree on overlapping sites. CDF-KL compares them separately and does not reconstruct a global distribution.

\subsection{Gaussian kernels and MC products}\label{app:mmd-training}
For binary coordinates,
\begin{equation}
e^{-(x_i-y_i)^2/(2\sigma^2)}=(1-\rho)+\rho(-1)^{x_i+y_i}.
\end{equation}
Multiplying over sites expands the kernel as
$k_\sigma(\mathbf{x},\mathbf{y})=\sum_Sw_\sigma(S)(-1)^{\sum_{i\in S}(x_i+y_i)}$.
Substitution into the kernel mean discrepancy gives Eq.~\eqref{eq:mmd}. Setting weights outside $\Omega$ to zero defines a positive-semidefinite restricted kernel. Distributions agreeing on the retained coefficients have zero restricted MMD even when they differ in omitted coefficients. This motivates the higher-order evaluation in Fig.~\ref{fig:transfer}.

With a fixed empirical target $\bar c_S$, use independent batches of sampled bitstrings in
\begin{equation}
\widehat L=\sum_{S\in\Omega}w_S
(\widehat c_S^{(1)}-\bar c_S)(\widehat c_S^{(2)}-\bar c_S).
\end{equation}
The two model estimates use independent draws, making $\widehat L$ unbiased for the restricted squared discrepancy to the fixed empirical target. Squaring a single noisy coefficient error would instead add its variance. Individual estimates can be negative. Finite target samples still contribute error.

In the Gaussian-sampled comparison in Table~\ref{tab:restriction-errors}, 162 nonempty supports of even size are drawn with replacement at each update, with probabilities proportional to the full Gaussian weights. The sample average is multiplied by their total kernel mass. Repeated supports count as repeated draws, so 162 draws do not imply 162 distinct coefficients. Odd-order model coefficients are zero. Their empirical target residuals add a parameter-independent term, which is omitted from the optimization. At $\sigma=1$, this sampling includes even orders above four. It therefore matches the coefficient-evaluation draw budget of the CDF study, not its exact information set.

The combined objective uses equal coefficients for CDF-KL and order-four MMD in the reported nine-site fits. The MMD weights are not renormalized after restricting orders. These conventions fix the relative contribution of the two losses.

\subsection{Bias-corrected MMD evaluation}\label{app:mmd-evaluation}
The sampled MMD in Figs.~\ref{fig:large} and~\ref{fig:large-2d} requires correcting both model and target squared coefficients. For independent, identically distributed real estimator draws $Y_1,\ldots,Y_N$ with mean $c_S$, the diagonal-subtraction construction used for IQP MMD~\cArmengol gives
\begin{equation}
\widehat{c_S^2}=\frac{(\sum_{j=1}^N Y_j)^2-\sum_{j=1}^N Y_j^2}{N(N-1)}
=\bar Y^2-\frac{s_Y^2}{N},\qquad N>1,
\end{equation}
where $s_Y^2$ is the sample variance with denominator $N-1$. This identity also applies to our Pauli coefficient estimator. If model and target coefficients use independent evaluation samples, their squared difference has the unbiased estimate
\begin{equation}
\widehat{(c_S(q)-c_S(p))^2}
=(\bar Y-\bar T)^2-\frac{s_Y^2}{N}-\frac{s_T^2}{M}.
\end{equation}
Here $T_1,\ldots,T_M$ are target parity samples, with $M>1$. Evaluation draws are independent of model fitting and selection. Averaging the corrected squared differences over supports drawn from the Gaussian kernel weights estimates Eq.~\eqref{eq:mmd}. Appendix~\ref{app:large} gives the sample budgets and grouping used for the error bars. The coefficient plots in Figs.~\ref{fig:fourier} and~\ref{fig:transfer} use exact probabilities and require no MC variance correction.

\section{Training, acquisition and local refinement}\label{app:training}
\subsection{Joint parameter updates across reference distributions}\label{app:joint-updates}
The trainable parameters $\etav=(\tht_0,\phiv)$ comprise the shared base circuit angles and the hypernetwork weights. For fixed reference data, let $\ell_j$ denote the loss at $\la_j$. A shared update differentiates the mean loss,
\begin{align}
\mathcal L(\etav)&=\frac1K\sum_{j=1}^K\ell_j(\tht_\etav(\la_j)),\\
\nabla_\etav\mathcal L&=\frac1K\sum_{j=1}^K
\left[\frac{\partial\tht_\etav(\la_j)}{\partial\etav}\right]^{\mathsf T}
\nabla_\tht\ell_j(\tht_\etav(\la_j)).
\label{eq:shared-update}
\end{align}
The hypernetwork takes $\la_j$ as input, while the corresponding reference samples or coefficients define $\ell_j$. In the small-system phase-map fits, all reference points form one batch. For larger systems, we accumulate their gradients sequentially before applying one optimizer update. In both cases, each update learns from several distributions. For MC objectives, the random inputs remain fixed during differentiation in Eq.~\eqref{eq:shared-update}.

For likelihood training, let $\widehat p_j$ be the empirical distribution of the reference samples at $\la_j$. Then
\begin{equation}
\mathcal L_{\rm NLL}=\frac1K\sum_{j=1}^K
\left[\KL(\widehat p_j\|q_{\etav,\la_j})+H(\widehat p_j)\right],
\label{eq:empirical-kl}
\end{equation}
where $H(\widehat p_j)$ is the Shannon entropy. This term is fixed by the data, so NLL and empirical KL give the same parameter gradients. The CDF-KL and MMD fits use the same shared update with their respective losses.

\subsection{Likelihood benchmarks and initialization}\label{app:initialization}
The small-system likelihood fits use full batches with equal reference-point weights and the adaptive moment estimation (Adam) optimizer~\cKingma. Circuit probabilities are obtained by summing over ancilla outcomes. We select checkpoints by empirical validation NLL. Three TFIM initializations share the same sampled datasets. The Haldane map comparison uses one initialization.

The RBM follows the construction of Ref.~\cTonner, with likelihood training using all outcome probabilities.

The TFIM hypernetwork IQP model uses all single- and two-site $X$ strings on nine qubits, giving 45 angles. Adding three ancillas gives 78 angles. Six ancilla-only rotations leave the measured distribution unchanged because they commute through the remaining IQP gates. Both models use GHZ preparation on the complete register. The conditional-rotation variants encode the field affinely through $R_y$ gates before the first Hadamard layer of the diagonal-$Z$ representation. These rotations can break the global-flip symmetry. We compare them with shared IQP angles and with hypernetwork-generated angles.

\begin{figure*}[!t]
\includegraphics[width=\textwidth]{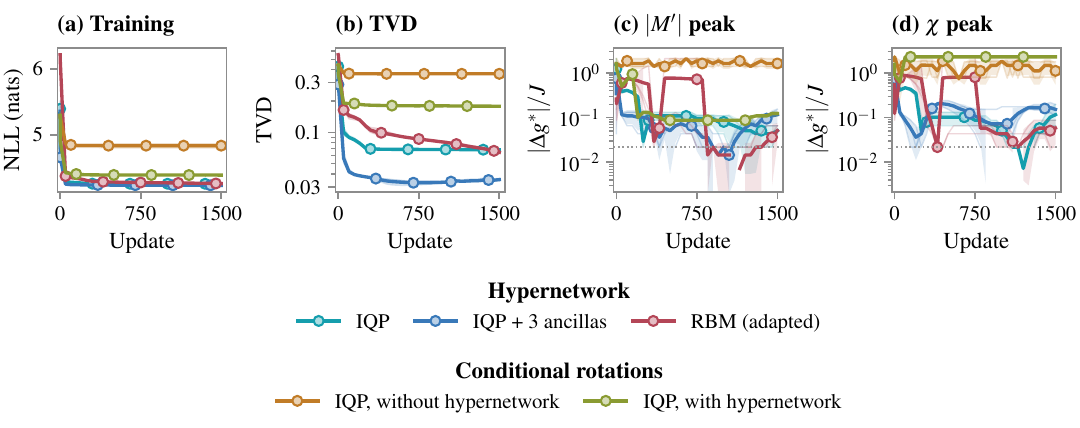}
\caption{\textbf{Training and response accuracy for the nine-site TFIM.} (a) Training NLL and (b) training-grid TVD for the five architectures in Fig.~\ref{fig:tfim}. (c,d) Absolute errors in the locations of the magnetic-response and susceptibility maxima. Thin curves show three diagnostic fits per model. Thick curves and bands show their mean and one sample standard deviation, with peak errors calculated before averaging. Dotted lines mark the field-grid spacing.}
\label{fig:learning}
\end{figure*}

The evaluation grid contains 161 fields, including 144 outside the training and validation sets. Figure~\ref{fig:learning} follows three diagnostic fits of each architecture using the benchmark settings. NLL and TVD are averaged equally over the nine training fields. Response peaks are the global maxima on the evaluation grid, including its endpoints, with ties resolved by choosing the smallest field.

In Fig.~\ref{fig:learning}, the three-ancilla IQP model reaches a training-grid TVD of about $0.035$, roughly half that of the RBM and IQP without ancillas. This agrees with the ordering of held-out infidelities in Fig.~\ref{fig:tfim}. Conditional rotations give higher TVD, with or without a hypernetwork. Peak-position errors continue to fluctuate after the training loss has largely flattened, and the ancilla model does not consistently give the closest peaks. Lower distribution error therefore need not imply a more accurate peak position.

\subsection{Adaptive reference selection}\label{app:acquisition}
Acquisition follows Eqs.~\eqref{eq:acquisition} and~\eqref{eq:acqstep}, adapted from Ref.~\cZhou. The small-system feature vector contains every nonconstant coefficient through order four. For larger systems, it combines the nonconstant coefficients of the 32 selected windows. We standardize each feature using its mean and standard deviation over acquired references, with the standard deviation floored at $0.05$.

The encoder has two tanh hidden layers of width 32 and an eight-dimensional latent vector. The decoder reconstructs the feature vector. At each acquisition stage, the autoencoder takes 200 Adam updates from its previous weights with fresh optimizer moments. The average $\mu_t$ includes all acquired references. We keep the model's MC draws fixed within each search stage when calculating gradients with respect to the Hamiltonian parameters.

A search starts from an acquired point and takes eight steps of size $0.1$ along the normalized gradient, with each Hamiltonian parameter rescaled to $[0,1]$. With probability $0.2$, we instead choose the point farthest from the acquired references among 2,048 uniformly drawn candidates. Accepted points must have Euclidean separation at least $0.006$ in these rescaled parameters. Small-system validation and evaluation points are excluded from acquisition. The 49-point protocol starts with a $3\times3$ grid and adds ten references in each of four rounds. Each of the five training stages takes 300 updates, weighting the available reference points equally.

We calculate reference states only at accepted points. The acquisition gradient passes through the model coefficients and encoder, so it requires no derivative of the reference solver.

\subsection{Combined-loss local refinement}\label{app:finetuning}
Figure~\ref{fig:transfer} uses CDF-KL pretraining and local CDF-KL plus MMD with equal loss weights. The local MMD kernel has bandwidth $\sigma=1$ and is truncated through order four. We estimate each of the 561 nonzero even-order coefficients with two independent groups of 32 MC draws. CDF-KL uses their mean and MMD uses their cross product. Independent fits use data-informed initialization.

At the 40 boundary points, we store the exact-diagonalization targets as untruncated MPSs and use 5,000 target samples for each local fit. For each initialization and pretraining method, we select a checkpoint by validation NLL. Angle-only and hypernetwork fine-tuning start from the same pretrained circuit angles. At each boundary point, we reset Adam moments and decrease the learning rate linearly from 0.003 to 0.00003 over 1,500 updates. Local evaluation uses the final update.

For the bandwidth evaluation, the Gaussian kernel is $K_\sigma(x,y)=\exp[-d_H(x,y)/(2\sigma^2)]$, where $d_H$ is Hamming distance. We sum over all 4,096 coefficients with weights $\rho_\sigma^{|S|}(1-\rho_\sigma)^{12-|S|}$, where $\rho_\sigma=(1-e^{-1/(2\sigma^2)})/2$. This evaluates the full kernel from exact probabilities, including coefficient orders absent from the training objective.

\begin{figure*}[t]
\centering
\includegraphics[width=\textwidth]{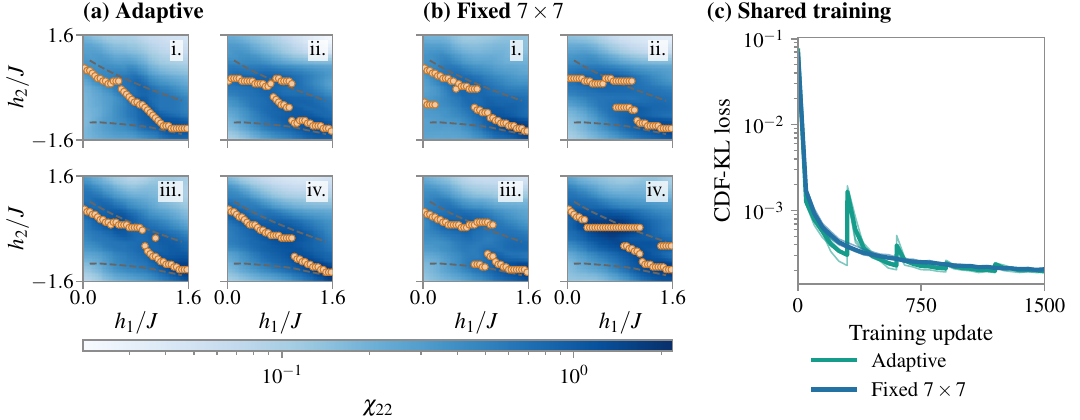}
\caption{\textbf{Pretraining the twelve-site models used in Fig.~\ref{fig:transfer}.} Susceptibility $\chi_{22}$ after (a) adaptive and (b) fixed-$7\times7$ training for four initializations, labeled i--iv. Orange circles mark learned response peaks and dashed curves show the bulk guides. (c) CDF-KL training loss averaged over the available references. Thin curves show individual initializations and thick curves their mean. Adaptive training grows from nine to 49 references, while fixed-grid training uses all 49 throughout.}
\label{fig:transfer-learning}
\end{figure*}
Figure~\ref{fig:transfer-learning} shows the four initializations used in Fig.~\ref{fig:transfer}, with panels i--iv corresponding to seeds 0, 1, 2 and 7. Both fixed-grid and adaptive training reach CDF-KL losses of about $2\times10^{-4}$. The adaptive loss rises when new references are added and then decreases within each stage. Despite similar final losses, the learned peak locations vary across initializations. The maps therefore provide information about the learned response beyond the training loss.

\subsection{CDF-KL-only comparison}\label{app:cdf-only}
Table~\ref{tab:transfer} uses CDF-KL alone for local training, in contrast to the combined loss in Fig.~\ref{fig:transfer}. It averages three initializations with the same data and MC draws. Independent QCBMs start from the angles produced by an untrained hypernetwork at each coordinate.

Fine-tuning uses 1,500 Adam updates at learning rate $3\times10^{-4}$ and optimizes both the hypernetwork and base circuit angles. Independent QCBMs use learning rate $0.003$. Higher-order coefficient errors are used only for evaluation.

Adaptive fine-tuning has the lowest mean coefficient error at every displayed order. The two fine-tuned models have similar errors through order six, but fine-tuning after adaptive pretraining gives lower errors at orders eight to twelve. At order twelve, its error is about four times smaller than that of independent training. The improvement therefore extends to coefficients absent from the local loss.

\begin{table}[t]
\centering\begin{tabular}{rrrrrr}\toprule
$k$ & A & F & Independent & A+FT & F+FT\\\midrule
2&7.21&5.95&0.243&0.224&0.230\\
4&2.75&3.01&1.67&1.27&1.27\\
6&1.29&1.53&0.892&0.681&0.682\\
8&0.762&1.04&0.615&0.436&0.508\\
10&0.577&1.07&0.625&0.361&0.623\\
12&0.522&2.04&1.52&0.387&1.16\\\bottomrule
\end{tabular}
\caption{\textbf{Coefficient errors after CDF-KL-only training for the twelve-site Haldane target.} Mean squared errors at order $k$, in units of $10^{-3}$, averaged over the same 40 boundary points and three initializations. A and F denote adaptive and fixed-grid pretraining, and FT denotes local fine-tuning.}
\label{tab:transfer}
\end{table}

\section{Large-system reference calculations and diagnostics}\label{app:large}
This appendix gives the reference calculations, training budgets and held-out evaluation used in Figs.~\ref{fig:large} and~\ref{fig:large-2d}. For both geometries, we sample DMRG reference states in the computational basis and use separate training and evaluation samples at the local fitting point.
\subsection{MPS references for the chain}
An MPS represents amplitudes as contractions of site tensors with finite bond dimensions~\cite{perez-garcia_matrix_2007,orus_practical_2014a}. We use Julia/ITensor DMRG~\cFishman. Its spin-$1/2$ operators require coefficients $8J$, $2h_1$ and $4h_2$ for the three-, one- and two-site terms to reproduce Eq.~\eqref{eq:haldane}. The chain has open boundaries.

Each chain reference calculation uses two initial states and retains the lower-energy result. The default bond-dimension cap is 128, the truncation cutoff is $10^{-10}$, and the total-energy stopping tolerance is $10^{-8}$. We allow three to twelve sweeps. The local eigensolver uses tolerance $10^{-12}$ and Krylov dimension ten. The stopping tolerance measures energy changes between sweeps.

For the same spin-$1/2$ cluster-Ising Hamiltonian, Ref.~\cCong located bulk phase boundaries using infinite-system DMRG with bond dimension at most 150. Its 135-site calculations used bond dimension 130, with no visible change in the reported figures upon increasing it to 150.

The chain references are optimized in the global-$X=+1$ sector, represented as a conserved parity after local Hadamards. We transform the MPS back to the physical computational basis before sampling.

Sequential MPS sampling generates outcomes directly from the retained state without Markov-chain equilibration~\cFerris. Each chain reference supplies 20,000 outcomes for the six-bit histograms and training coefficients. The sampled distribution retains the MPS approximation, which can be more restrictive near critical regions~\cTagliacozzo.

At the evaluation coordinate $(h_1/J,h_2/J)=(1.46,-1.40)$, the MMD comparison uses held-out samples from the same cap-128 MPS used to generate the training data.

\subsection{Sparse circuit and response evaluation}\label{app:large-circuit}
Both geometries use the same circuit on 1,024 measured qubits and 342 unmeasured ancillas. We use single-site $X$ rotations, nearest-neighbor $XX$ rotations in circuit order and paired $Y/Z$ rotations. These gates give $r=2$, so each of the two evolved vectors has four amplitudes. Choosing gates to keep $r$ small requires care, since some choices also yield circuits that are easy to sample classically.

CDF-KL uses 32 disjoint six-site windows. Each marginal has 63 nonconstant coefficients, of which 31 have even order. For MMD, we draw Fourier supports over all 1,024 measured qubits, with 683 draws per bandwidth and update at $\sigma\simeq 3.82,\,8.35,\,11.2$. We average the squared MMD over these bandwidths and add it to CDF-KL with weight $\alpha_{\rm MMD}=0.893$ for the chain and $0.905$ for the lattice, rounded to three significant figures. These weights are calibrated using nine references with 5,000 samples each. Shared training and both local fits use the same loss within each geometry.

For both geometries, training uses 20,000 target bitstrings per reference and 1,024 MC draws per estimated model coefficient. The training set starts with nine points and adds 8, 8, 8, 8 and 9 points in successive stages. The first five stages take 200 updates each and the final stage takes 500. Each update averages gradients over the available reference points and uses fresh Fourier supports and model MC draws. We evaluate the final model after 1,500 updates.

After exact gate cancellation, each observable calculation uses a proper subset of the circuit sites. Uniform even-parity bitstrings on the full register have a uniform marginal on any such subset, since an omitted bit can complete either parity. We can therefore draw uniform bitstrings on the remaining sites while preserving the GHZ expectation.

For the response maps in Figs.~\ref{fig:large}(a) and~\ref{fig:large-2d}(a), we evaluate the 992 nonconstant even-order coefficients on a $40\times40$ grid, using 1,000 MC draws per coefficient at each point. We keep these draws fixed across the grid and differentiate through the coefficient estimates, marginal projection and smoothing to obtain the selected-marginal susceptibility in Eq.~\eqref{eq:localchi}.

\subsection{Local refinement and held-out evaluation}
The local evaluation points are $(h_1/J,h_2/J)=(1.46,-1.40)$ for the chain and $(1.22,-0.04)$ for the lattice. At each point, both local fits use the same 20,000 bitstrings from the reference MPS and take 600 Adam updates. Fine-tuning starts from the final shared model and updates its base angles and hypernetwork at learning rate $0.0003$. Independent training optimizes the circuit angles at learning rate $0.003$.

For panel (b) of each figure, we draw one million held-out bitstrings from the MPS that supplied the local training samples. We divide them into 40 groups of 25,000 for the chain and 20 groups of 50,000 for the lattice. Each bandwidth uses 400,000 Gaussian-weighted Fourier support draws in total. Each lattice group has 20,000 draws. The chain assigns 10,016 draws to half the groups and 9,984 to the rest, preserving batches of 32 supports. Each sampled model coefficient uses 100,000 auxiliary MC draws, shared within each batch and across the three models. Squared coefficients use distinct pairs of draws, and the cross term uses independent model and target samples.

The black curve compares the 20,000 local training samples with the held-out samples by calculating the Gaussian kernel directly on the bitstrings. We split the training sample into groups of 500 for the chain and 1,000 for the lattice, pair each with its corresponding held-out group and exclude self-pairs from the within-sample sums. All curves show the unweighted mean over groups. Error bars give the standard error, the sample standard deviation across groups divided by the square root of the number of groups. They measure evaluation uncertainty for fixed trained models.

\subsection{Rectangular-lattice training and evaluation}
Figure~\ref{fig:large-2d} uses a 4 by 256 rectangular lattice with open boundaries. The Hamiltonian contains $-h_1X_i$ at each site and $-h_2X_iX_j$ on each nearest-neighbor edge. Sites are ordered along the four-site direction first. For each edge $(i,j)$ with $i<j$, the Hamiltonian also contains $-JZ_iX_jZ_k$ for every neighbor $k$ of $j$ at graph distance two from $i$. We use this same site ordering for the MPS, whose contraction cost depends on entanglement and the range of interactions~\cOrus.

Each reference uses two initial states in the global-$X=+1$ sector and retains the lower-energy DMRG result. The bond cap is 128, with four to twelve sweeps, cutoff $10^{-8}$, total-energy stopping tolerance $10^{-5}$ and local eigensolver tolerance $10^{-8}$. We transform back to the physical computational basis before sampling.

The lattice uses the same circuit and training schedule as the chain. Its 32 six-site marginals are $2\times3$ rectangles distributed along the long direction of the lattice. The largest detected response peak on the displayed map is at $(1.26,0.04)$.

\section{Additional numerical comparisons}\label{app:additional}
\subsection{Comparisons of loss and gate restrictions}
Table~\ref{tab:restriction-errors} extends the comparison in Fig.~\ref{fig:restrictions} to additional training objectives. All infidelities are evaluated from exact model probabilities on the same nine-site grid. Gaussian-sampled MMD gives a small reduction in mean infidelity relative to order-four MMD, while the combined CDF-KL and MMD loss gives a larger error. The Gaussian-sampled objective includes higher even orders at a matched number of coefficient draws. Exact and MC training give similar errors for each objective.

\begin{figure*}[!t]
\includegraphics[width=\textwidth]{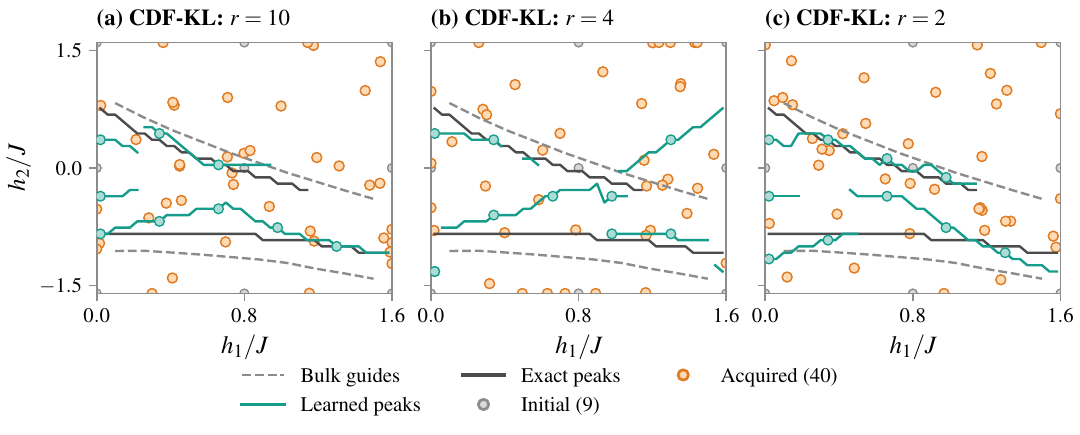}
\caption{\textbf{Haldane susceptibility peaks with different gate restrictions.} Adaptive CDF-KL fits use (a) $r=10$, (b) $r=4$ and (c) $r=2$, with the same angle count, MC budget and order-four loss. Learned peaks are compared with exact nine-site peaks and the bulk guides~\cCong. Points show each fit's nine initial and 40 acquired references.}
\label{fig:ranks}
\end{figure*}

Figure~\ref{fig:ranks} compares the learned susceptibility peaks for three gate restrictions. All three fits recover parts of the two response branches. The $r=4$ fit develops an upper branch that rises away from the reference at larger $h_1/J$, while the $r=2$ fit follows the descending reference branch more closely. The two restricted fits have nearly equal mean infidelity, about $0.107$, despite these differences in their peaks. Each fit selects its own acquisition points, so the comparison includes changes in both the circuit and the selected data.

\begin{table*}[t]
\centering\begin{tabular}{llrrlr}\toprule
Objective & Point selection & Points & $r$ & Coefficients & Mean $\mathcal I(p,q)$\\\midrule
Likelihood & Fixed grid &100&10&Full probabilities&0.00670\\
Likelihood & Fixed grid &49&10&Full probabilities&0.0175\\
CDF-KL & Fixed grid &49&10&Exact, $k\leq4$&0.0743\\
CDF-KL & Fixed grid &49&10&MC, $k\leq4$&0.0751\\
MMD & Fixed grid &49&10&Exact, $k\leq4$&0.0497\\
MMD & Fixed grid &49&10&MC, $k\leq4$&0.0496\\
CDF-KL & Adaptive &49&10&MC, $k\leq4$&0.0819\\
MMD & Adaptive &49&10&MC, $k\leq4$&0.0655\\
CDF-KL & Adaptive &49&4&MC, $k\leq4$&0.107\\
CDF-KL & Adaptive &49&2&MC, $k\leq4$&0.107\\
MMD & Adaptive &49&4&MC, $k\leq4$&0.0841\\
MMD & Adaptive &49&2&MC, $k\leq4$&0.0906\\
Gaussian-sampled MMD & Fixed grid &49&10&Exact, 162 draws&0.0483\\
Gaussian-sampled MMD & Fixed grid &49&10&MC, 162 draws&0.0484\\
CDF-KL + MMD & Fixed grid &49&10&Exact, $k\leq4$&0.0561\\
CDF-KL + MMD & Fixed grid &49&10&MC, $k\leq4$&0.0549\\\bottomrule
\end{tabular}
\caption{\textbf{Mean Haldane distribution infidelity on the $40\times40$ evaluation grid.} Each row is one fit. Exact and MC specify coefficient evaluation during training, with 64 MC draws per nonzero coefficient.}
\label{tab:restriction-errors}
\end{table*}

\vspace{-12pt}
\subsection{Other nine-site targets}
\vspace{-6pt}
We apply the fixed $10\times10$ protocol of Fig.~\ref{fig:haldane} to the alternating Heisenberg chain, the axial next-nearest-neighbor Ising (ANNNI) chain, and a cluster chain with Ising interactions and a transverse field. These Hamiltonian families also appear in the phase-classification benchmarks of Ref.~\cite{bermejo_quantum_2026}. A separate hypernetwork is trained for each target. With $X_j,Y_j,Z_j$ denoting Pauli operators on site $j$, the Hamiltonians are
\begin{align}
H_{\rm Heis}&=\frac14\sum_{j=1}^{n-1}J_j(X_jX_{j+1}+Y_jY_{j+1}+Z_jZ_{j+1}),\label{eq:target-heis}\\
H_{\rm ANNNI}&=-\sum_{j=1}^{n-1}X_jX_{j+1}+\kappa\sum_{j=1}^{n-2}X_jX_{j+2}-h\sum_{j=1}^nZ_j,\label{eq:target-annni}\\
H_{\rm C}&=\sum_{j=1}^n(Z_j-J_1X_jX_{j+1}-J_2X_jZ_{j+1}X_{j+2}).\label{eq:target-cluster}
\end{align}
In Eq.~\eqref{eq:target-heis}, $J_j=J_{\rm odd}$ on odd bonds and $J_j=J_{\rm even}$ on even bonds. Both couplings are antiferromagnetic, and the factor $1/4$ converts Pauli products to spin-$1/2$ products. In Eq.~\eqref{eq:target-annni}, the nearest-neighbor ferromagnetic coupling sets the energy unit, $\kappa\geq0$ is the competing antiferromagnetic next-nearest-neighbor coupling, and $h$ is the transverse field in the same units. In Eq.~\eqref{eq:target-cluster}, the transverse-field coefficient is fixed to $+1$, $J_1$ controls the two-site Ising interaction, and $J_2$ controls the three-site cluster interaction $X_jZ_{j+1}X_{j+2}$. The ANNNI parameters $\kappa,h$ and the cluster parameters $J_1,J_2$ are dimensionless.

Heisenberg and ANNNI have open boundaries, while cluster indices are taken modulo $n$. The parameter domains are $(J_{\rm odd},J_{\rm even})\in[0.25,1]^2$, $(\kappa,h)\in[0,1.5]\times[0,2]$ and $(J_1,J_2)\in[-2,2]^2$. The Heisenberg reference is selected in the global-$X=+1$ sector. At degeneracies we use a fixed seeded ground-state selection and omit undefined responses.

Dashed gray curves in Figs.~\ref{fig:heis-extra}--\ref{fig:cluster-restrictions} show bulk phase boundaries, distinct from the nine-site susceptibility peaks. For the alternating Heisenberg chain, the bulk critical line is $J_{\rm even}=J_{\rm odd}$. The ANNNI guides follow the phase diagram used in Refs.~\cite{bermejo_quantum_2026,cea_exploring_2026}. We use the approximate Ising boundary
\begin{equation}
h_{\rm I}(\kappa)\simeq\frac{2-4\kappa}{1+\sqrt{(1-3\kappa+4\kappa^2)/(1-\kappa)}},
\label{eq:annni-bulk-ising}
\end{equation}
for $0\leq\kappa\leq1/2$, and the fitted Kosterlitz--Thouless (KT) and Pokrovsky--Talapov (PT) boundaries for $\kappa>1/2$~\cite{beccaria_evidence_2007},
\begin{align}
h_{\rm KT}(\kappa)&\simeq1.05\sqrt{(\kappa-\tfrac12)(\kappa-0.1)},\label{eq:annni-bulk-kt}\\
h_{\rm PT}(\kappa)&\simeq1.05(\kappa-\tfrac12).\label{eq:annni-bulk-pt}
\end{align}
The KT and PT curves bound the floating phase above and below, respectively.

For the cluster Hamiltonian, the bulk single-particle excitation energy can be written as $\varepsilon(k)=2|1+J_1e^{ik}+J_2e^{2ik}|$, where $k$ is crystal momentum and the lattice spacing is one. Its zeros give
\begin{equation}
J_2=-1\pm J_1,\qquad J_2=1\ \text{with}\ |J_1|\leq2.
\label{eq:cluster-bulk}
\end{equation}
These are the boundaries used in Ref.~\cite{bermejo_quantum_2026}, consistent with the free-fermion solution after accounting for the transverse-field sign~\cite{niu_majorana_2012}.

For each target, the architecture comparison uses 100 training points with 5,000 samples each, 24 separate validation points and a $40\times40$ evaluation grid. Each model uses one initialization, 1,500 likelihood updates and checkpoint selection by validation likelihood. Model budgets and ancilla locations follow Fig.~\ref{fig:haldane}. Circuit preparation is GHZ for Heisenberg and the computational zero state for ANNNI and cluster. The RBM enforces bit-complement symmetry, which generally does not hold for the ANNNI and cluster target distributions in this basis. All susceptibilities are evaluated from exact model probabilities, with derivatives taken along the vertical parameter axis.

The restriction studies use Haldane QCBMs for all three targets and follow Fig.~\ref{fig:restrictions}. They compare 100- and 49-point likelihood fits, exact and MC coefficient evaluation, fixed and adaptive 49-point training, and gates restricted to $r=2$. The coefficient losses use orders through four, retaining only even orders for Heisenberg and odd and even orders for ANNNI and cluster. All fits use 1,500 updates, with 64 MC draws per nontrivial coefficient when required. The circuits retain three unmeasured ancillas, 298 angles and 9,581 hypernetwork weights. Each adaptive fit selects 40 new references from nine initial points.

\subsubsection{Alternating Heisenberg chain}
\begin{figure*}[p]
\centering
\includegraphics[width=0.68\textwidth,height=0.40\textheight,keepaspectratio]{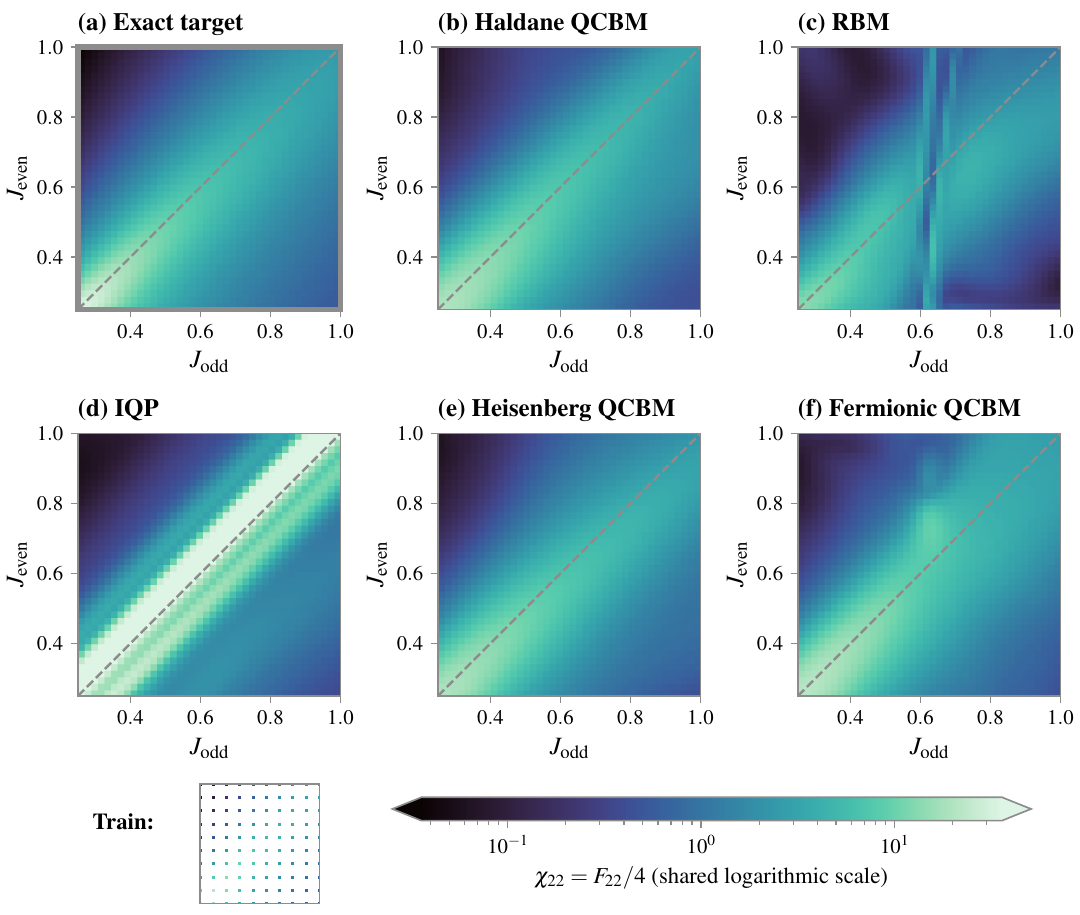}
\caption{\textbf{Learned susceptibility for the nine-site open alternating Heisenberg chain.} Panels show (a) the exact target, (b) Haldane QCBM, (c) RBM, (d) IQP, (e) Heisenberg QCBM and (f) Fermionic QCBM. The response is $\chi_{22}$ with respect to $J_{\rm even}$. The dashed line marks $J_{\rm even}=J_{\rm odd}$, and the inset shows the training grid.}
\label{fig:heis-extra}
\par\medskip
\includegraphics[width=0.68\textwidth,height=0.35\textheight,keepaspectratio]{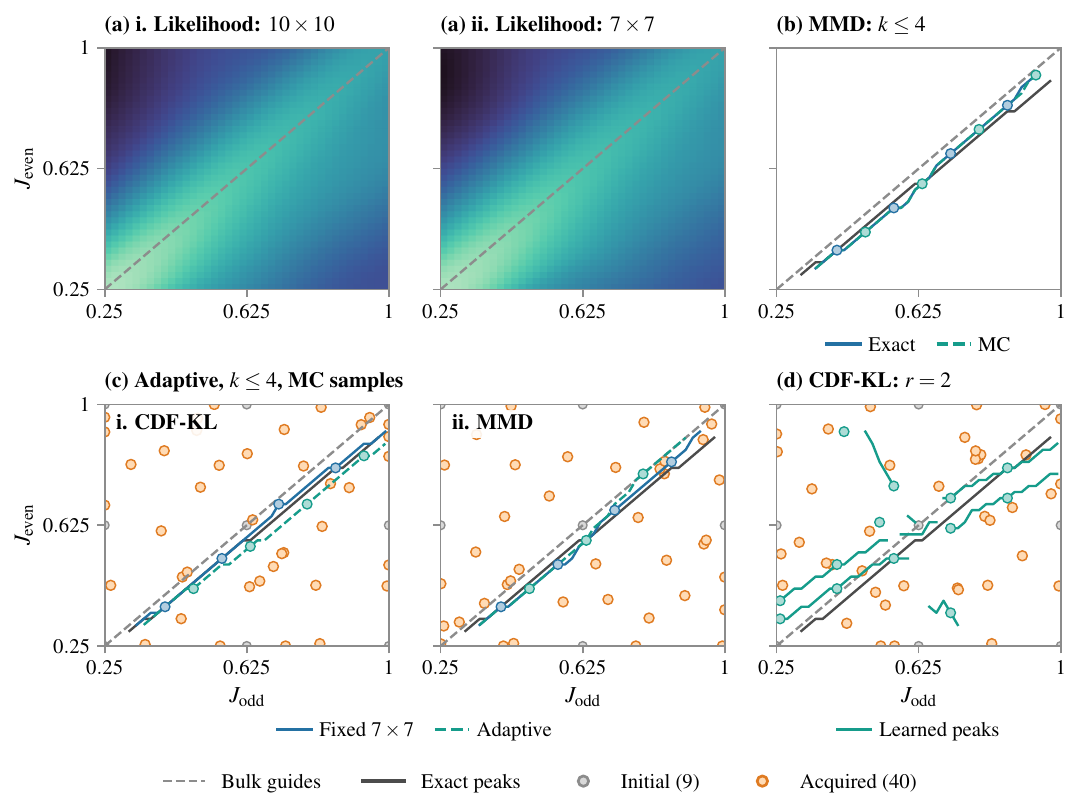}
\caption{\textbf{Restricted training for the alternating Heisenberg target.} Haldane QCBMs use (a.i,ii) likelihood on $10\times10$ and $7\times7$ grids, (b) order-four MMD with exact or MC coefficients, (c.i,ii) fixed or adaptive CDF-KL and MMD, and (d) adaptive CDF-KL with $r=2$. Curves compare learned and exact nine-site susceptibility peaks with the bulk guide. Points mark initial and acquired references.}
\label{fig:heis-restrictions}
\end{figure*}

In Fig.~\ref{fig:heis-extra}, the Haldane and Heisenberg QCBMs closely reproduce the broad susceptibility band of the exact target. Both have mean infidelity about $0.0020$, compared with $0.0154$ for the Fermionic QCBM, $0.158$ for IQP and $0.208$ for the RBM. IQP places a much sharper band near the diagonal, while the RBM introduces a pronounced vertical feature absent from the target.

Figure~\ref{fig:heis-restrictions} shows that reducing the likelihood grid from 100 to 49 points leaves the response map nearly unchanged. Order-four training also follows the main sequence of finite-size peaks, with close agreement between exact and MC MMD fits. Fixed and adaptive CDF-KL have similar mean infidelities, $0.0202$ and $0.0208$. The additional gate restriction to $r=2$ produces extra peak branches and increases the error to $0.310$.

\subsubsection{ANNNI chain}
\begin{figure*}[p]
\centering
\includegraphics[width=0.68\textwidth,height=0.40\textheight,keepaspectratio]{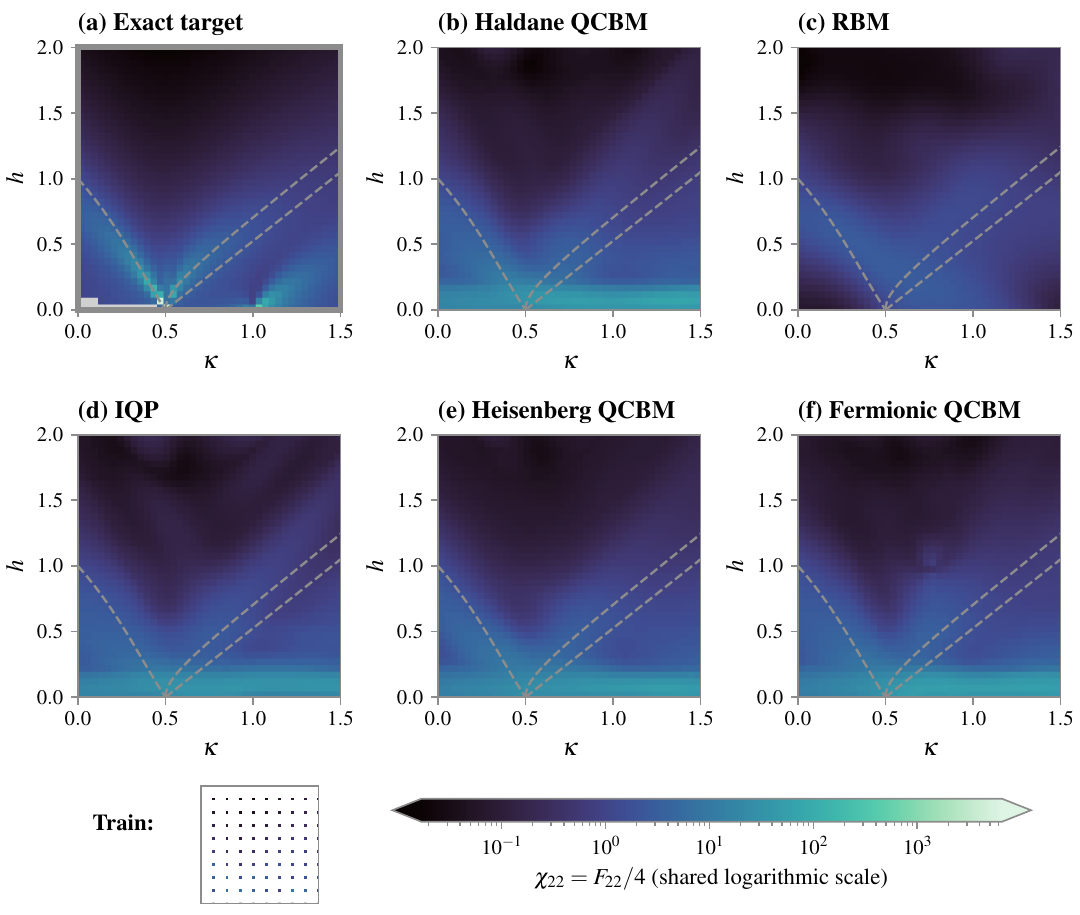}
\caption{\textbf{Learned susceptibility for the nine-site open ANNNI chain.} Panels follow Fig.~\ref{fig:heis-extra}, with the response taken with respect to $h$. The inset shows the training grid. Dashed curves are the approximate bulk boundaries in Eqs.~\eqref{eq:annni-bulk-ising}--\eqref{eq:annni-bulk-pt}.}
\label{fig:annni-extra}
\par\medskip
\includegraphics[width=0.68\textwidth,height=0.34\textheight,keepaspectratio]{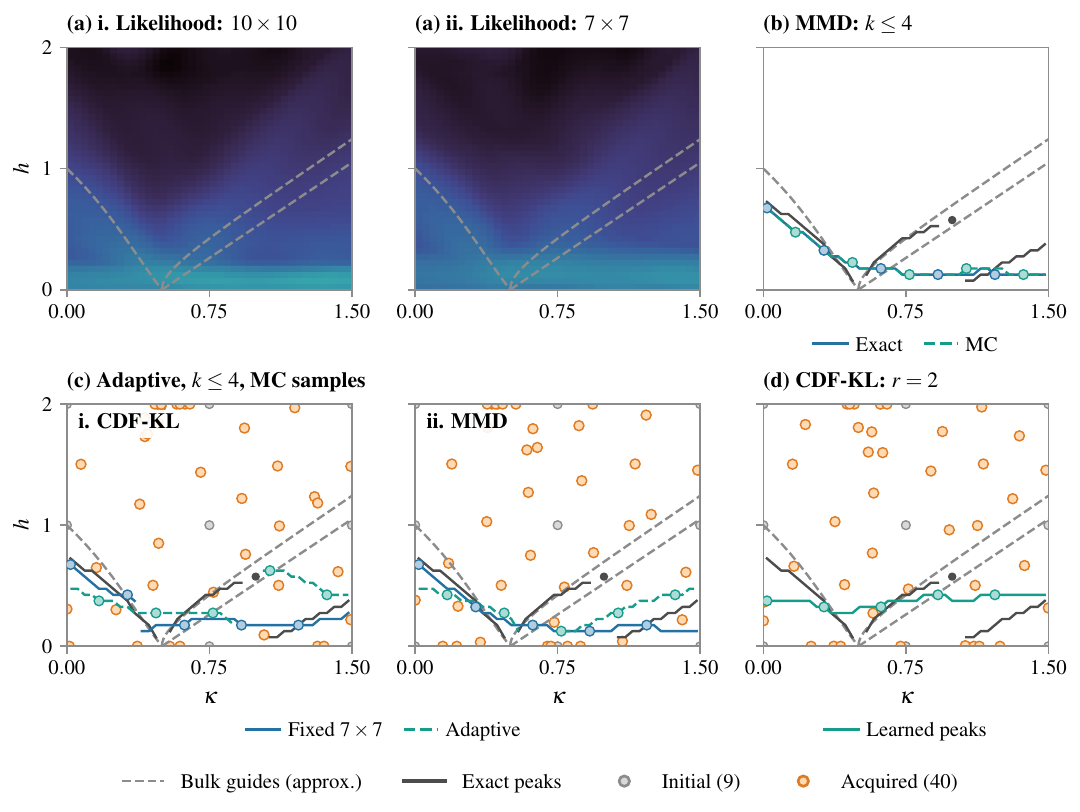}
\caption[Restricted ANNNI training]{\textbf{Restricted training for the ANNNI target.} Panels and reference markers follow Fig.~\ref{fig:heis-restrictions}. The response is taken with respect to $h$, and coefficient losses include orders one through four. Learned peaks are compared with exact nine-site peaks and the approximate bulk boundaries.}
\label{fig:annni-restrictions}
\end{figure*}

The four circuit models in Fig.~\ref{fig:annni-extra} give similar mean infidelities, between $0.0423$ and $0.0431$. They reproduce the broad response at larger fields, but replace the target's sharper low-field features by a nearly horizontal band. The RBM error is much larger, $0.641$, with the symmetry restriction described above.

In Fig.~\ref{fig:annni-restrictions}, exact and MC MMD again give nearly identical peak locations. The fitted peaks follow the descending reference branch at small $\kappa$. Beyond $\kappa=0.5$, the fits recover only parts of the reference structure, and gaps in the two branches persist after adaptive training. Reducing the likelihood grid raises mean infidelity from $0.0430$ to $0.0586$, while order-four MMD gives $0.221$. Here the loss restriction has a larger effect than replacing exact coefficients by MC estimates.

\subsubsection{Periodic cluster chain}
\begin{figure*}[p]
\centering
\includegraphics[width=0.68\textwidth,height=0.40\textheight,keepaspectratio]{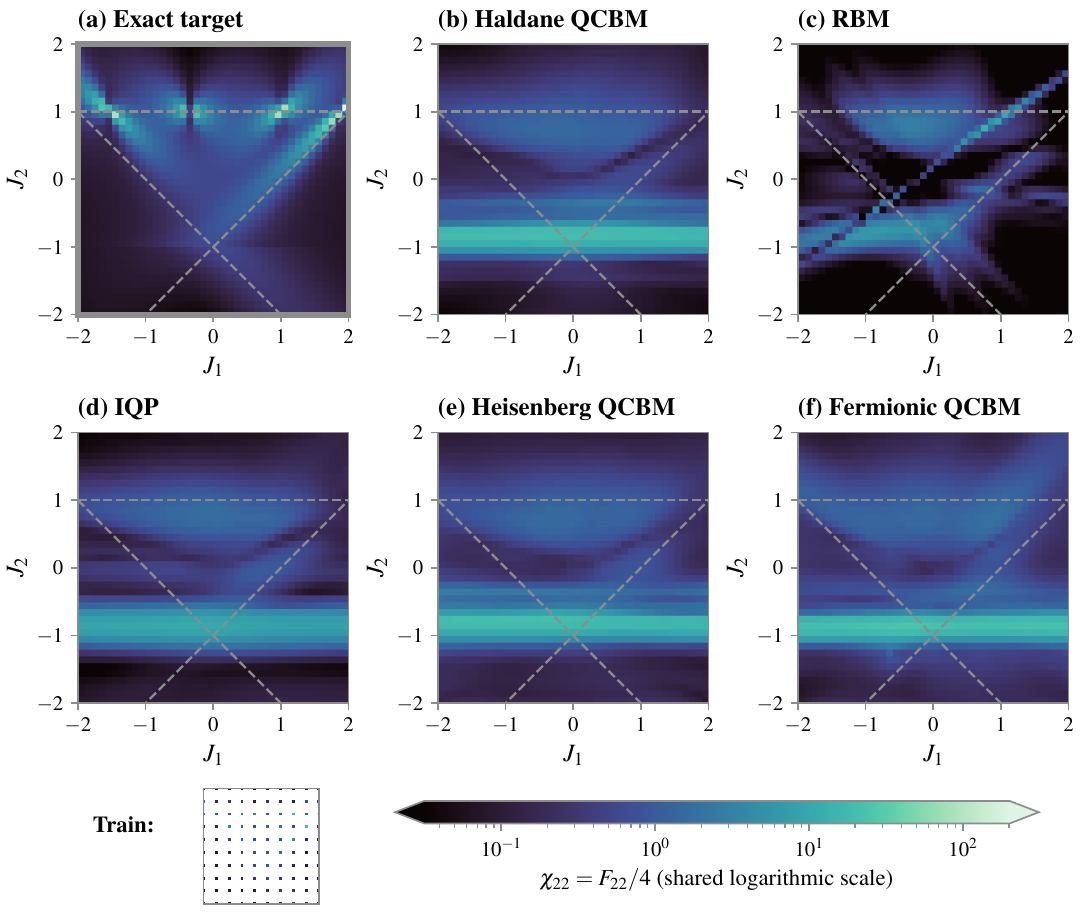}
\caption{\textbf{Learned susceptibility for the nine-site periodic cluster chain.} Panels follow Fig.~\ref{fig:heis-extra}, with the response taken with respect to $J_2$. The inset shows the training grid. Dashed lines mark the bulk boundaries in Eq.~\eqref{eq:cluster-bulk}.}
\label{fig:cluster-extra}
\par\medskip
\includegraphics[width=0.68\textwidth,height=0.35\textheight,keepaspectratio]{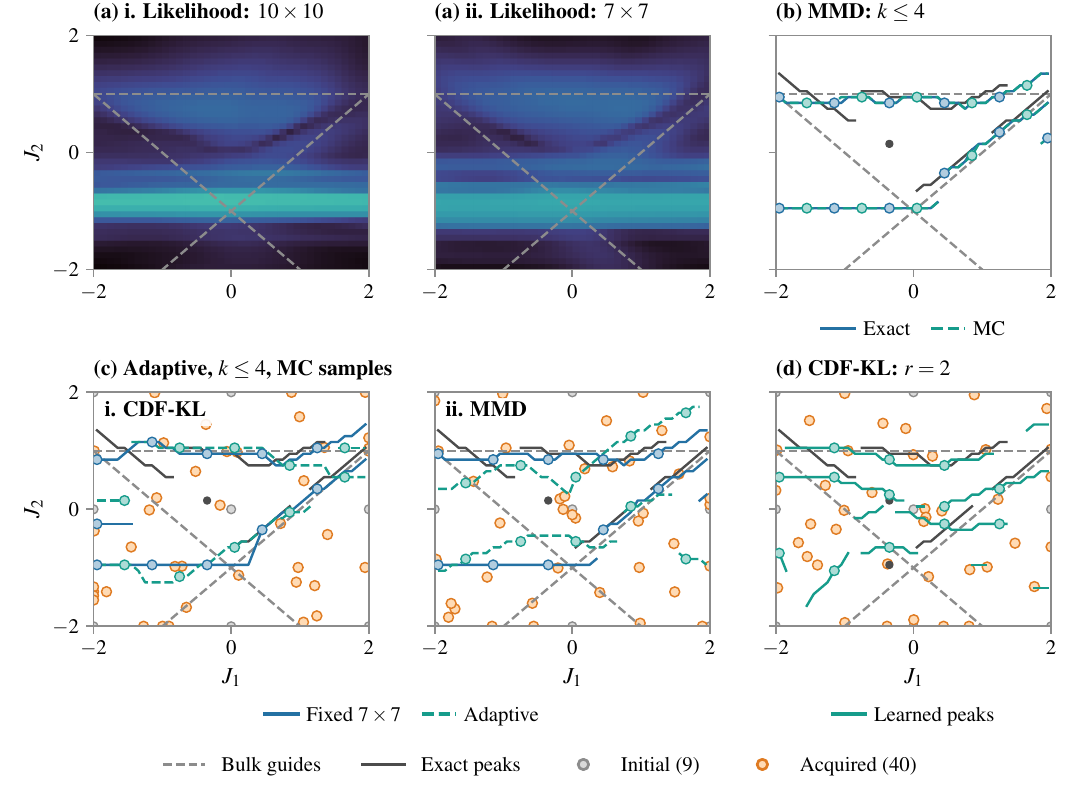}
\caption[Restricted cluster training]{\textbf{Restricted training for the periodic cluster target.} Panels and reference markers follow Fig.~\ref{fig:heis-restrictions}. The response is taken with respect to $J_2$, and coefficient losses include orders one through four. Learned peaks are compared with exact nine-site peaks and the bulk boundaries.}
\label{fig:cluster-restrictions}
\end{figure*}

For the cluster target in Fig.~\ref{fig:cluster-extra}, the Haldane, Heisenberg and Fermionic QCBMs have mean infidelities between $0.0823$ and $0.0903$, compared with $0.154$ for IQP and $0.663$ for the symmetry-restricted RBM. The circuit maps resolve the broad response near $J_2=1$. The exact target has additional narrow features along the upper branch, while a horizontal band near $J_2=-0.8$ appears only in the circuit maps.

Figure~\ref{fig:cluster-restrictions} shows little change between the two likelihood grids. Order-four MMD recovers parts of the upper sequence of peaks and the ascending branch at positive $J_1$, with similar results from exact and MC coefficients. Its mean infidelity is nevertheless about $0.437$, compared with $0.0882$ for the 49-point likelihood fit. Adaptive and restricted-gate fits show larger changes in peak locations.


\end{document}